\documentclass[article]{jfm}

\usepackage{graphicx}
\usepackage{epstopdf,epsfig}
\usepackage[]{subcaption} 
\usepackage{amsmath}
\usepackage{enumitem} 
\usepackage{placeins}
\usepackage{newtxtext}
\usepackage{newtxmath}
\usepackage{natbib}
\usepackage{hyperref}
\hypersetup{
    colorlinks = true,
    urlcolor   = blue,
    citecolor  = blue,
}
\newcommand{\aRA}{\ensuremath{\langle a \rangle}}
\newcommand{\RomanNumeralCaps}[1]
\linenumbers

\title{Power-generation enhancements and upstream flow properties of turbines in unsteady inflow conditions}

\author{Nathaniel J. Wei\aff{1}
 \and John O. Dabiri\aff{1}\aff{2}
 \corresp{\email{jodabiri@caltech.edu}}}

\affiliation{\aff{1}Graduate Aerospace Laboratories, California Institute of Technology, Pasadena, CA 91125, USA
\aff{2}Department of Mechanical and Civil Engineering, California Institute of Technology, Pasadena, CA 91125, USA}

\begin{document}
\maketitle

\begin{abstract}
Energy-harvesting systems in complex flow environments, such as floating offshore wind turbines, tidal turbines, and ground-fixed turbines in axial gusts, encounter unsteady streamwise flow conditions that affect their power generation and structural loads. In some cases, enhancements in time-averaged power generation above the steady-flow operating point are observed. To characterize these dynamics, a nonlinear dynamical model for the rotation rate and power extraction of a periodically surging turbine is derived and connected to two potential-flow representations of the induction zone upstream of the turbine. The model predictions for the time-averaged power extraction of the turbine and the upstream flow velocity and pressure are compared against data from experiments conducted with a surging-turbine apparatus in an open-circuit wind tunnel at a diameter-based Reynolds number of $Re_D = 6.3\times10^5$ and surge-velocity amplitudes of up to 24\% of the wind speed. The combined modeling approach captures trends in both the time-averaged power extraction and the fluctuations in upstream flow quantities, while relying only on data from steady-flow measurements. The sensitivity of the observed increases in time-averaged power to steady-flow turbine characteristics is established, thus clarifying the conditions under which these enhancements are possible. Finally, the influence of unsteady fluid mechanics on time-averaged power extraction is explored analytically. The theoretical framework and experimental validation provide a cohesive modeling approach that can drive the design, control, and optimization of turbines in unsteady flow conditions, as well as inform the development of novel energy-harvesting systems that can leverage unsteady flows for large increases in power-generation capacities.
\end{abstract}



\section{Introduction}
\label{sec:intro}

Energy-harvesting turbines in atmospheric and oceanic flows are routinely exposed to unsteady flow conditions from gusts, tides, turbulent fluctuations, and other strongly time-dependent fluid motions. These unsteady effects induce time-varying forces and loads on the turbine components, which impact the time-averaged efficiency and operational lifespans of these systems. Oscillations in the streamwise velocity incident on the turbine are thus of major concern for conventional ground-fixed turbines in axial gusts, hydrokinetic turbines in tidal flows, and turbines mounted to airborne kites, which may undergo rapid changes in incident wind speed as they sweep through atmospheric flow gradients. These dynamic-inflow conditions are also related by means of a reference-frame transformation to the problem of a turbine moving in periodic linear surge motions in a steady inflow. This problem is of particular interest for emerging offshore-wind technologies, such as floating offshore wind turbines (FOWTs). Since these turbines are not fixed to the ocean floor, they can move under the influence of wind gusts and surface waves. Of these motions, the linear-surge oscillation mode tends to exhibit larger amplitudes relative to other degrees of freedom of motion \citep{johlas_large_2019}. In certain forcing scenarios and platform configurations the velocity amplitude of the turbine motions may exceed 25\% of the wind speed \citep{wayman_coupled_2006,larsen_method_2007,de_vaal_effect_2014}. In spite of these complicating factors, FOWTs have the potential to enable wind-energy conversion in areas of the ocean whose depths prohibit the installation of conventional fixed-bottom systems, thereby creating additional avenues for the expansion of wind power as a contributor to global energy demands. They can capitalize on strong offshore wind resources, are by nature located close to coastal urban centers, and have fewer constraints on size and placement compared to their land-based and seafloor-mounted counterparts. The characterization of the aerodynamics of oscillating turbines, in addition to that of stationary turbines in oscillatory inflow conditions, is thus of critical importance to the design and control of the next generation of wind-energy technologies.

Unsteady streamwise flow conditions are particularly intriguing from both a fluid-mechanics and engineering perspective because they have the potential to yield substantial increases in the time-averaged power extraction of an energy-harvesting system. The one-dimensional (1D) axial momentum theory developed by \cite{betz_maximum_1920} (as well as Lanchester, Joukowski, and others) posits that the power-conversion efficiency of an energy-harvesting system may not exceed $C_{p,Betz}=16/27\approx59.3\%$, but this analysis was conducted under the assumption of steady flow. \cite{dabiri_theoretical_2020} recently relaxed that assumption and suggested that the contribution of an unsteady velocity potential could lead to theoretical efficiencies in excess of the so-called Betz limit. In parallel with this prediction, several studies have shown relative power enhancements over the steady operating power for turbines in surge motions or unsteady flows, both experimentally \citep{farrugia_investigating_2014,el_makdah_influence_2019,mancini_characterization_2020,wei_phase-averaged_2022} and in simulations of varying fidelity \citep{farrugia_study_2016,wen_influences_2017,johlas_floating_2021}. However, the extent to which unsteady flow physics contributed to these observed power enhancements is unclear. For example, \cite{wen_influences_2017} and \cite{johlas_floating_2021} found that a quasi-steady model for the time-averaged power could qualitatively describe the trends in the power enhancements. \cite{mancini_characterization_2020}, by contrast, found that data from wind-tunnel experiments exceeded the predictions of their quasi-steady solution, though this solution differed from that of \cite{johlas_floating_2021}. Additionally, both \cite{farrugia_study_2016} and \cite{wei_phase-averaged_2022} found that the magnitude of the power enhancements depends on the turbine tip-speed ratio, and that under certain conditions, time-averaged power losses relative to the steady case are also possible. A full explanation and parameterization of these divergent observations remains lacking in the literature, and is urgently needed if the floating offshore wind-turbine technologies currently under development are to take advantage of these effects for increased power-generation capabilities.

The 1D axial momentum theory of Betz also asserts that the deceleration of the upstream flow approaching a wind turbine, or induction, is coupled with the operation and power output of a turbine. This induction effect dictates the flow and loading conditions encountered by the blades of a turbine and is directly related to the turbine's thrust force and efficiency. The induction zone, defined roughly as the region in which the flow velocity along the turbine's centerline is below 95\% of the free-stream velocity, extends at least two turbine diameters upstream of the turbine itself \citep{medici_upstream_2011}. These reduced velocities can thus bias tower-based estimates of the true wind speed by anemometers and LiDAR systems \cite[e.g.][]{larsen_full-scale_2014,howard_upwind_2016,simley_characterization_2016,borraccino_wind_2017,mann_how_2018}. For floating turbines, the coupling between incident wind conditions, blade-pitch control systems, and turbine thrust can also yield negatively damped (i.e.\ unstable) streamwise surge oscillations that increase fatigue loading on the turbine blades \citep{larsen_method_2007,jonkman_influence_2008,lopez-queija_review_2022}. It is therefore instructive for turbine modeling and control to quantify the coupling between unsteady streamwise flow conditions, the dynamics of the turbine, and the flow properties in the upstream induction region.

The flow deceleration upstream of a stationary horizontal-axis wind turbine has been thoroughly studied in the literature, and several parameterizations of the induction region exist. One frequently used modeling approach treats the wake of the turbine as a cylindrical vortex sheet \citep{branlard_cylindrical_2015}. This model lends itself well to free-vortex wake simulations \citep{sarmast_comparison_2016}, and shows good agreement with experimental data \citep{medici_upstream_2011,howard_upwind_2016,bastankhah_wind_2017,borraccino_wind_2017}. It has also been extended to unsteady inflow conditions \citep{chattot_actuator_2014,yu_new_2019}. Rather than rely on assumptions regarding near-wake structure, alternative approaches model the induction effect of the turbine using potential-flow objects such as Rankine half-bodies \citep{araya_low-order_2014,gribben_potential_2019,meyer_forsting_verification_2021} or porous discs \citep{modarresi_flow_1979}. These models reflect the common practice in both numerical and experimental studies of modeling the turbine as an actuator disc. Other models, such as the self-similar solution of \cite{troldborg_simple_2017}, are better able to capture the radial dependence of the streamwise flow velocity in the induction region. To the authors' knowledge, these models have not yet been extended to dynamically varying streamwise inflow conditions, such as axial gusts or turbine surge motions, as most existing studies involving these flow conditions do not investigate the upstream induction region.

The lack of parameterizations for the time-averaged power enhancements of turbines in unsteady inflow conditions and their coupled upstream flow properties motivates the present theoretical and experimental study. The work is structured as follows. First, in Section \ref{sec:model}, a nonlinear dynamical model for the power extraction of a periodically surging turbine is derived, and a method is proposed that couples the time-varying power generated by the turbine to the turbine induction. This modeling framework is combined with two induction models to yield time-resolved predictions of the flow field upstream of the surging turbine. These predictions rely solely on turbine data obtained from steady-flow measurements, namely the turbine power curve and the streamwise velocity averaged radially across the face of the rotor. A brief note on the dynamic equivalence between a surging turbine in a steady inflow and a stationary turbine in an oscillatory inflow is also presented. In Section \ref{sec:methods}, velocity and pressure measurements upstream of a surging-turbine apparatus are described, and the time-averaged power extraction and flow measurements are compared with the predictions of the modeling framework in Section \ref{sec:results}. Additional analyses of the sensitivity of the modeling framework to the steady-flow aerodynamics of the turbine and the role of unsteady fluid mechanics are presented in Section \ref{sec:discussion}. Finally, implications of the findings for the design, optimization, and control of turbines in unsteady flow environments are discussed.

\section{Nonlinear Dynamics of a Periodically Surging Turbine}
\label{sec:model}

In this section, we derive a nonlinear dynamical model for the power extraction and flow properties upstream of a periodically surging turbine. We present a nonlinear ordinary differential equation for the turbine rotation rate as a function of known steady-flow turbine-aerodynamics and generator characteristics. This model can predict the time-varying and time-averaged rotation rate, torque, and power of the turbine. By applying 1D momentum theory, the axial induction factor of the turbine can be estimated from the instantaneous turbine power, and coupling this estimate with flow models allows the flow velocity and pressure at any point upstream of the turbine to be predicted. The modeling framework captures the unsteady dynamics of the surging-turbine problem using a quasi-steady parameterization of the turbine aerodynamics; potential contributions from unsteady fluid dynamics are explored later in Section \ref{sec:unsteady}.

In our notation, time-averaged quantities are marked with overbars, steady-flow or quasi-steady quantities are labeled with a zero subscript (e.g.\ $\mathcal{P}_0$), spatial averages are denoted with angle brackets, and amplitudes are denoted with a circumflex (e.g.\ $\hat{u}$). Additionally, if a flow variable lacks a specified radial dependence, it refers to a quantity measured on the turbine centerline (i.e.\ $r=0$).

\subsection{A nonlinear model for turbine rotation rate and power extraction}
\label{sec:model_pwr}

We build upon the linear modeling approach of \cite{wei_phase-averaged_2022}, who describe the time-varying dynamics of a turbine using the swing equation (i.e.\ Newton's second law for rotation),

\begin{equation}
    J \frac{d\omega}{dt} = \tau_{aero} - \tau_{gen},
    \label{eqn:swing}
\end{equation}

\noindent where $J$ is the moment of inertia of the turbine system about its axis of rotation and $\omega$ is the rotation rate of the turbine. Deviations from equilibrium between the aerodynamic and generator torques $\tau_{aero}$ and $\tau_{gen}$ produce changes in the turbine rotation rate. \cite{wei_phase-averaged_2022} model the generator torque using the ordinary differential equation for the torque from a permanent-magnet motor,

\begin{equation}
    \tau_{gen} = K_2 \frac{d\omega}{dt} + K_1 \omega + K_0,
    \label{eqn:tau_gen}
\end{equation}

\noindent where $K_2$ is the moment of inertia of the generator about its axis of rotation, $K_1$ is the generator constant, and $K_0$ is an empirical offset. These parameters can be established empirically for a given generator over a range of resistive loads.

While \cite{wei_phase-averaged_2022} utilized a linearized model for the aerodynamic torque, in this work we explicitly include the nonlinear relationship between the turbine coefficient of power,

\begin{equation}
C_p = \frac{\mathcal{P}}{\frac{1}{2}\rho \pi R^2 u_\infty^3},
\end{equation}

\noindent and the tip-speed ratio,

\begin{equation}
\lambda = \frac{R\omega}{u_\infty},
\end{equation}

\noindent where $\mathcal{P}$ is the power extracted by the turbine from the flow, $\rho$ is the fluid density, $R$ is the radius of the turbine, and $u_\infty$ is the free-stream flow velocity. Any given turbine has a power curve defined as $C_p = C_{p,0}(\lambda)$, which has a local maximum at a power-maximizing tip-speed ratio $\lambda_{opt}$. Note that in this work, we will use the subscript $0$ to refer to steady-flow quantities. Since the turbine power is determined by the product of the torque on the turbine and its rotation rate, the torque on the turbine can be written in terms of the power curve as

\begin{equation}
\tau = \frac{1}{2}\rho \pi R^2 u_\infty^3 \frac{C_{p,0}(\lambda)}{\omega}.
\end{equation}

For a given surge-velocity profile $U(t)$, the effective free-stream velocity is $u_\infty = u_1 - U(t)$, where we define $u_1$ as the far-field wind speed relative to a ground-fixed frame. Thus, we may write the aerodynamic torque as

\begin{equation}
    \tau_{aero} = \tau_{aero}(\omega,U,t) = \frac{1}{2}\rho \pi R^2 \frac{\left(u_1-U(t)\right)^3}{\omega} C_{p,0}\left(\frac{R\omega}{u_1-U(t)}\right).
    \label{eqn:tau_aero}
\end{equation}

Substituting Equations \ref{eqn:tau_gen} and \ref{eqn:tau_aero} into Equation \ref{eqn:swing} results in a nonlinear ordinary differential equation for the turbine rotation rate that is first-order in time and depends on the surge velocity as an input forcing parameter:

\begin{equation}
    \frac{d\omega}{dt} = \frac{1}{J+K_2}\left[\frac{1}{2}\rho \pi R^2 \frac{\left(u_1-U(t)\right)^3}{\omega} C_{p,0}\left(\frac{R\omega}{u_1-U(t)}\right) - K_1 \omega - K_0\right].
    \label{eqn:model_rotRate}
\end{equation}

This model is a nonlinear and nonautonomous ordinary differential equation, which precludes straightforward mathematical characterization, but it can be integrated forward in time from an initial condition $\omega(t=0)$ until it reaches a period-averaged equilibrium. The model can therefore yield numerical predictions of the time-varying and time-averaged rotation rate, torque, and power of a turbine under surge motions or dynamic inflow conditions. This stands in contrast to the linearized model developed by \cite{wei_phase-averaged_2022}, which can be written as a transfer function for convenient analysis but is unable to capture changes in time-averaged quantities.

As a limiting case, we may consider a quasi-steady solution to the model in which $\frac{d\omega}{dt}=0$ and $C_{p,0}$ is constant as a function of time. The time-averaged power is defined as $\overline{\mathcal{P}} = \overline{\tau_{gen}\omega} = K_1 \overline{\omega^2} + K_0 \overline{\omega}$, which suggests that for a sinusoidal surge-velocity waveform with amplitude $u^* = \hat{U}/u_1$, the time-averaged power is
\begin{equation}
    \overline{\mathcal{P}} = \frac{f}{2\pi} \int_0^{\frac{2\pi}{f}} \frac{1}{2}\rho \pi R^2 C_{p,0} \left(u_1 - U(t)\right)^3 dt = \mathcal{P}_0\left(1+\frac{3}{2}{u^*}^2\right).
    \label{eqn:powerWenJohlas}
\end{equation}

This result is identical to that derived by \cite{wen_influences_2017} and \cite{johlas_floating_2021} for a surging turbine, and is equivalent to that of a stationary turbine with constant $C_p$ in an oscillating inflow.

\subsection{Modeling the relationship between turbine dynamics and upstream flow conditions}
\label{sec:model_upstream}

\begin{figure}
\centerline{\includegraphics[width=1.0\textwidth]{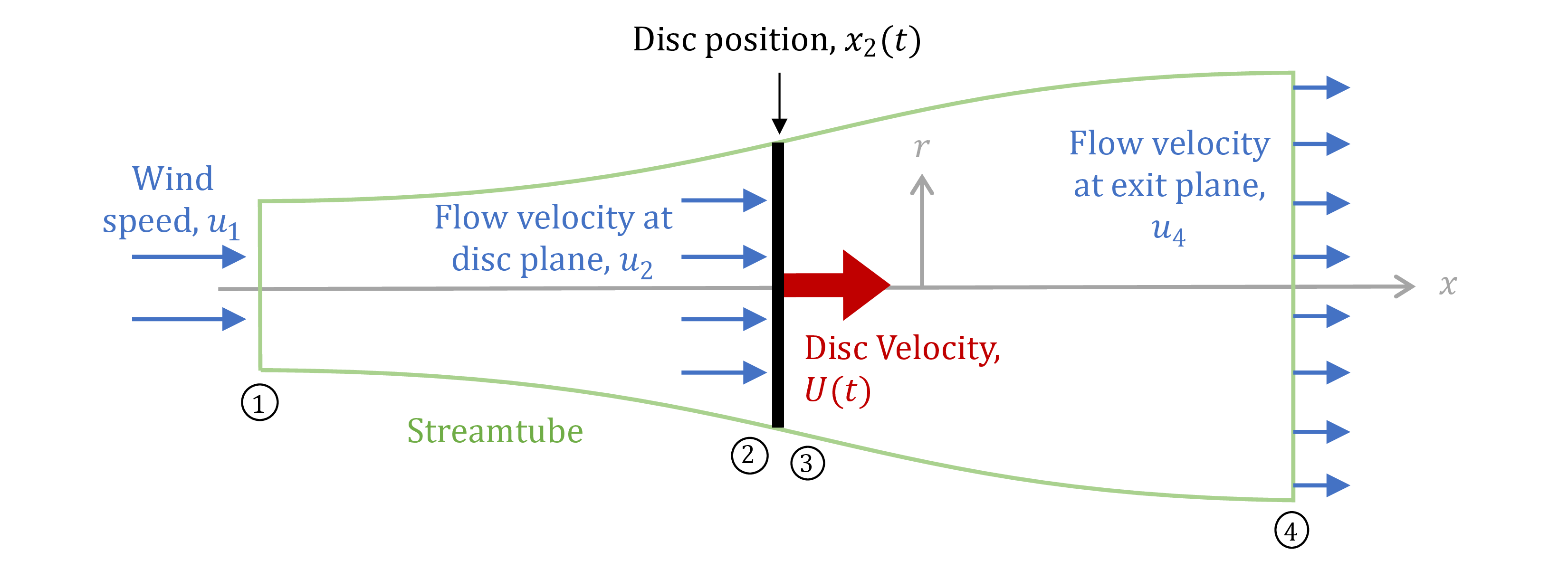}}
\caption{Schematic of the parameters and control volumes referenced in Section \ref{sec:model}. The actuator disc is located instantaneously at $x_2(t)$ and moves with velocity $U(t)$ relative to the inertial frame defined by the $x$- and $r$-axes. Circled numbers denote streamwise interrogation locations (1 through 4).}
  \label{fig:CV}
\end{figure}

The 1D momentum theory derived by \cite{betz_maximum_1920} can be used to infer flow properties upstream of the turbine rotor plane from the power extracted by the turbine. This theoretical framework employs conservation relations over a control volume composed of an axisymmetric streamtube surrounding the turbine, which is modeled as an actuator disc. An axial induction factor, representing the decrease in velocity from far upstream of the turbine to the upstream face of the actuator disc (i.e.\ location 2 in Figure \ref{fig:CV}), is defined as

\begin{equation}
    a = \frac{u_1 - u_2}{u_1}.
    \label{eqn:indFactor}
\end{equation}

\noindent The induction factor is the single free parameter needed to compute the coefficient of power within this framework, which is given by 

\begin{equation}
    C_p = 4a(1-a)^2.
    \label{eqn:CpBetz}
\end{equation}

\noindent This yields a theoretical maximum for the efficiency of a wind-energy system of $C_{p,Betz} = 16/27$, which is attained at $a=1/3$. For the surging-turbine system, then, we have the similar relation
\begin{equation}
    \frac{\tau_{gen}\omega}{\frac{1}{2}\rho \pi R^2 u_1^3} = 4a(1-a)^2.
    \label{eqn:CpSurge}
\end{equation}

1D momentum theory assumes that the flow is inviscid, incompressible, irrotational, and steady. Upstream of the turbine rotor plane, the first three assumptions are reasonable to make. If we further assume that the flow upstream of the turbine can be modeled in a quasi-steady manner, we can use the rotation rate given by Equation \ref{eqn:model_rotRate} to compute an instantaneous coefficient of power $C_p(\omega,t)$, and by inverting Equation \ref{eqn:CpSurge}, we can estimate the time-varying induction factor $a(t)$. Since the induction factor is physically constrained as $a\in[0,1]$, two solutions are possible for $C_p < C_{p,Betz}$. The upper solution for $a$ represents a heavily loaded turbine, and for $a\gtrsim0.37$ the theoretical framework breaks down \citep{wilson_applied_1974}. In a majority of cases, including the experiments presented in this work, the turbine is not heavily loaded, and thus the lower solution for $a$ is assumed to represent the system.

Equations \ref{eqn:model_rotRate} and \ref{eqn:CpSurge} therefore connect the time-varying dynamics of a turbine under dynamic axial-flow conditions to the flow properties just upstream of the rotor disc. To propagate these predictions further upstream, an induction model for the turbine is needed. As mentioned previously, a common modeling approach involves representing the wake of the turbine as a cylindrical vortex sheet and performing Biot-Savart integration to compute the induced velocity from this wake model at any point in the domain \citep{johnson_helicopter_1980,branlard_cylindrical_2015}. This is known as vortex-sheet or vortex-cylinder theory (hereafter abbreviated as VCT). Evaluating this integral upstream of the turbine along its rotational axis yields a model for the induced velocity along the upstream centerline of the turbine \citep{medici_upstream_2011}:
\begin{equation}
\frac{u(r=0,x)}{u_\infty} = 1 - a\left[1+\frac{x-x_2}{R}\left(1+\left(\frac{x-x_2}{R}\right)^2\right)^{-1/2}\right],
\label{eqn:vortextheory}
\end{equation}
where $x$ is the streamwise coordinate along the axis of the turbine (originating at the turbine and positive downstream) and $x_2$ is the instantaneous location of the turbine, as shown in Figure \ref{fig:CV}. For a surging turbine, the induction effect in the second term should scale with the effective free-stream velocity $u_1-U(t)$, since in the limiting case where the turbine is moving downstream at velocity $U(t) = u_1$, the turbine should have no effect on the flow, and the flow everywhere upstream of the turbine should be equal to $u_1$. Involving the time-varying induction factor $a(t)$ provided by 1D momentum theory, we thus obtain the flow velocity at any upstream location $x<x_2$ and at any time as
\begin{equation}
    u(r=0,x,t) = u_1 - a(t)(u_1-U(t))\left[1+\frac{x-x_2(t)}{R}\left(1+\left(\frac{x-x_2(t)}{R}\right)^2\right)^{-1/2}\right].
\label{eqn:VCT}
\end{equation}

To complete the description of flow properties upstream of the surging turbine, the pressure along the centerline may be modeled by substituting the model for $u(r=0,x,t)$ into the steady Bernoulli equation,
\begin{equation}
p(r=0,x) = p_1 + \frac{1}{2}\rho \left({u_1}^2 - {u\left(r=0,x,t\right)}^2\right),
\label{eqn:pres}
\end{equation}
where $\rho$ is the density of the fluid and $p_1$ is the ambient pressure in the free stream. If the velocity potential of the induction model is known, the unsteady Bernoulli equation may be applied instead. For the purposes of this work, however, we maintain quasi-steady assumptions for the flow physics in the induction zone, in keeping with the quasi-steady aerodynamics parameterized by the turbine model and 1D momentum theory. As mentioned previously, the effect of this unsteady potential will be considered in more detail in Section \ref{sec:unsteady}. While the expressions presented here have been confined to the centerline, \cite{branlard_cylindrical_2015} provide full relations for $u(r,x)$ that can be employed in place of Equation \ref{eqn:vortextheory} to allow this modeling framework to cover the entire upstream induction zone.

Alternatively, we can model the effect of the surging turbine on the flow using a porous-disc representation in potential flow, which does not rely on parameterizations of the turbine wake geometry and vorticity. This is inspired by the work of \cite{taylor_aerodynamics_1944} and \cite{koo_fluid_1973}, which has recently been extended by \cite{steiros_drag_2018} for flat plates of arbitrary porosity and \cite{bempedelis_analytical_2022} for wind turbines at arbitrary loading conditions. The porous-disc approach, which has not been widely investigated in the wind-energy literature, is presented to demonstrate the robustness of the overall modeling framework put forth in this work to the choice of induction model. Its velocity potential is also more readily accessible than that of VCT, which will be advantageous when we consider the effects of unsteady fluid mechanics in Section \ref{sec:unsteady}. Finally, the model provides a convenient generalization to rotors with arbitrary radial distributions of streamwise velocity, which, though not explored in detail in this study, could be exploited to integrate this model with blade-element momentum (BEM) computations that generate radially varying induction-factor profiles. We will refer to the model throughout this paper as the porous-disc theory (PDT).

\subsection{A porous-disc induction model for a surging turbine}
\label{sec:model_PDT}

We first consider a circular porous disc with radius $R$ located at streamwise coordinate $\xi=0$, represented as a distribution of sources with a velocity potential of $\phi(r,\xi=0) = C\sqrt{R^2-r^2}$ for $r<R$ and $\phi(r,\xi=0)=0$ for $r>R$, where $C>0$ is an arbitrary constant that represents the aggregate strength of the source distribution. Using this distribution as a boundary condition at $\xi=0$, we may solve the Laplace equation $\nabla^2\phi=0$ in cylindrical coordinates to obtain the velocity potential of a porous disc \cite[cf.][]{lamb_hydrodynamics_1916,tranter_bessel_1968}:
\begin{equation}
\phi\left(r,\xi\right) = -\sqrt{\frac{\pi}{2}} C R^{3/2} \int_0^\infty s^{-1/2}J_{3/2}(Rs)J_0(rs)e^{-s \xi} ds;\; \xi>0.
\label{eqn:phiPrime}
\end{equation}
Here, $J_\nu(z)$ is a Bessel function of the first kind, and $s$ is a dummy integration variable. The choice of $C=\frac{2}{\pi}V$ gives the velocity potential of a solid disc moving at axial velocity $V$ in a quiescent fluid \citep[\textsection 102.4]{lamb_hydrodynamics_1916}. More generally, the velocity $V$ represents the velocity of the disc relative to that of the fluid in the far field. For a porous disc, we may define a representative source term $a$, directly corresponding to the induction factor defined in Equation \ref{eqn:indFactor}, such that $C=\frac{2}{\pi}a V$. The choice of $a$ dictates the porosity of the disc: $a=0$ represents a fully permeable disc, $a=1$ yields a fully solid disc, and intermediate values ($0<a<1$) reduce the source strength from the solid-disc solution so that a nonzero mass flux through the disc is established. Evaluating Equation \ref{eqn:phiPrime} along the centerline yields
\begin{equation}
\phi\left(r=0,\xi\right) = -a V \frac{2}{\pi}\left[R - \xi\arctan\left(\frac{R}{\xi}\right)\right];\; \xi>0.
\label{eqn:phir0}
\end{equation}
This solution is only valid for $\xi>0$. To describe the other half of the domain as well, one might follow the ansatz of \cite{taylor_aerodynamics_1944} and use the even extension of $\phi$ to represent $\xi<0$. The velocity discontinuity across the disc that this extension creates could then be removed using the base-suction correction of \cite{steiros_drag_2018}. However, since in this work we are only concerned with the upstream region, we leave these derivations for future consideration.

We differentiate the velocity potential with respect to $\xi$ to obtain the streamwise velocity along the centerline:
\begin{equation}
u\left(r=0,\xi\right) = -a V \frac{2}{\pi}\left[\frac{R\xi}{{\xi}^2+R^2}-\arctan\left(\frac{R}{\xi}\right)\right];\; \xi > 0.
\label{eqn:uPrime}
\end{equation}
This relation emphasizes the effect of the porosity parameter (or equivalently, the induction factor) on the behavior of the model. For $a=0$, the flow is everywhere unaffected by the motion of the disc. For $a=1$, $u(r=0,\xi\rightarrow0^+)=V$, which satisfies the surface boundary condition for a moving solid disc.

We now apply this expression in an inertial frame containing a uniform flow with free-stream velocity $u_1$, in which the disc translates at velocity $U(t)$ relative to the frame. In this frame, we define the downstream-oriented axial coordinate $x$ and the instantaneous position of the disc $x_2(t)$ as shown in Figure \ref{fig:CV}, such that $\xi=x_2 - x$ and $U(t) = \frac{dx_2}{dt}$. The velocity of the disc relative to the far-field flow velocity is thus $V = U(t) - u_1$. Applying these definitions to Equation \ref{eqn:uPrime}, we arrive at the following model for the centerline velocity in the upstream induction zone ($x<x_2$):
\begin{equation}
u\left(r=0,x,t\right) = u_1 - a(t) \left(u_1-U(t)\right) \frac{2}{\pi}\left[\frac{R(x-x_2(t))}{\left(x-x_2(t)\right)^2+R^2}-\arctan\left(\frac{R}{x-x_2(t)}\right)\right].
\label{eqn:PDT}
\end{equation}
We reiterate that the model cannot be used to predict the velocity downstream of the porous disc, given the constraint of $\xi>0$ on the velocity potential. Additionally, though the solution is technically valid if the turbine moves downstream faster than the wind speed, i.e.\ $u_1-U(t) < 0$, we expect that the model will cease to be valid in this case because the rotor will interact with its own wake.

As with the expression obtained from VCT, Equation \ref{eqn:phiPrime} can in principle be integrated at any point upstream of the porous disc so that the velocity and pressure can be computed throughout the entire induction zone. We note that, in both models, the streamwise velocity $u_2$ on the upstream face of the rotor plane is predicted to be constant over $r$, i.e.\ 
\begin{equation}
u(x=x_2,r)=u_1-a(t)(u_1-U(t)).
\label{eqn:u2_r}
\end{equation}
This is generally a poor approximation for real turbines, whose streamwise velocities typically increase toward the free-stream value with increasing radial distance from the hub \cite[cf.][]{medici_upstream_2011,troldborg_simple_2017}. An additional advantage of the porous-disc modeling approach is that arbitrary radial-velocity profiles can be modeled by changing the source-strength distribution in Equation \ref{eqn:phiPrime}. For example, if the turbine blade geometry is known, radial variations in the induction factor can be computed from blade-element momentum theory as a function $a(r)$ and integrated to obtain a modified velocity potential. A correction factor for the effects of nonuniform velocities at the disc face can then be derived by defining a rotor-averaged induction factor,
\begin{equation}
\aRA = \frac{2}{R^2}\int_0^R a(r) r dr = \left[\frac{2}{R^2}\int_0^R \frac{a(r)}{a(r=0)} r dr\right] a(r=0) \equiv \kappa a(r=0).
    \label{eqn:aRotorAvg}
\end{equation}
The scaling constant $\kappa$, which maps the centerline induction factor to the equivalent rotor-averaged induction factor, can be computed from a known velocity-deficit radial profile, such as the self-similar solution of \cite{troldborg_simple_2017} or a BEM computation. For a top-hat velocity-deficit profile, $\kappa=1$. If the radial induction-factor distribution does not change much throughout the surge cycle (i.e.\ the velocity-deficit profile remains self-similar), $\kappa$ can be assumed to be independent of the surge kinematics for a given loading condition, and thus can be treated as a constant for that particular wind speed and mean tip-speed ratio.

The distinction between the centerline and rotor-averaged induction factors is particularly important for a quantitative comparison between the 2D axisymmetric induction theories developed in this section and the 1D axial-momentum theory of Betz. Because 1D momentum theory by definition does not account for radial differences in streamwise velocity, the induction factor estimated using Equation \ref{eqn:CpSurge} must represent the rotor-averaged induction factor. By the same logic, the induction factor in the VCT and PDT expressions for the streamwise velocity is the centerline induction factor, and is what will be measured by a point anemometer placed along the rotational axis of the turbine. The parameter $\kappa$ thus serves as a bridge between the 1D and 2D analyses employed in this modeling framework.

For the remainder of this paper, we will focus on measurements taken along the rotational axis of the turbine ($r=0$), and therefore the flow quantities $u(x)$, $p(x)$, and $a$ will implicitly refer to these centerline quantities.

The modeling approach presented in this work may be summarized as follows:

\noindent\hangindent=12pt 1) Identify the turbine power curve $C_{p,0}(\lambda)$ and generator and inertia constants $K_0$, $K_1$, $K_2$, and $J$ from steady-flow measurements and manufacturer specifications.

\noindent\hangindent=12pt 2) Integrate Equation \ref{eqn:model_rotRate} in time with a given surge-velocity forcing $U(t)$ to obtain the turbine rotation rate $\omega(t)$.

\noindent\hangindent=12pt 3) From $\omega(t)$, use Equations \ref{eqn:tau_gen} and \ref{eqn:CpSurge} to predict the coefficient of power $C_p(t)$ and the rotor-averaged induction factor $\langle a(t) \rangle$.

\noindent\hangindent=12pt 4) Calculate the centerline induction factor $a(t)$ using an analytical or empirical correction factor $\kappa$.

\noindent\hangindent=12pt 5) Include $a(t)$ in an induction model (e.g.\ Equation \ref{eqn:VCT} for VCT or \ref{eqn:PDT} for PDT) to obtain the velocity field upstream of the turbine.

\noindent\hangindent=12pt 6) Use the steady Bernoulli equation to obtain the pressure field from the velocity field.

We reiterate that this modeling framework invokes a quasi-steady assumption for the aerodynamics of the turbine, and the time dependence of the model comes from an unsteady treatment of the turbine rotation rate as a function of the aerodynamic and generator torques. The practical benefit of this approach is that time-resolved predictions of the turbine dynamics and upstream flow properties can be obtained solely on the basis of steady-flow measurements; the model does not depend on empirical calibrations from unsteady test cases. From a fluid-mechanics perspective, the approach provides an instructive disambiguation between the rotational dynamics of the turbine and the actuator-disc aerodynamics associated with the rotor and its motions. Furthermore, this analytical foundation allows the effects of unsteady flow physics to be more directly characterized.

\subsection{On the problem of a stationary turbine in an oscillating inflow}
\label{sec:model_oscInflow}

The preceding analysis has focused on the case of a periodically surging turbine in a uniform inflow. From the work of \cite{wen_influences_2017}, \cite{el_makdah_influence_2019}, \cite{johlas_floating_2021}, and others, it is apparent that the case of a stationary turbine in an axial gust can be made equivalent to the surging-turbine case by shifting from a ground-fixed to a turbine-fixed frame of reference. The time-averaged power is not affected by this transformation. To determine the effect of noninertial-frame accelerations, we consider the force on a body oscillating with velocity $W_i(t)$ in an oscillating inflow $U_i(t)$, which is given by \cite{brennen_review_1982} as
\begin{equation}
F_i = -M_{ij} \frac{dW_j}{dt} + \left(M_{ij} + \rho V_D \delta_{ij}\right) \frac{dU_j}{dt};\; j = 1,2,3.
\label{eqn:forceAMplusBuoyancy}
\end{equation}
Here, $M_{ij}$ is the added-mass tensor of the body, $V_D$ is the volume of the body, $\delta_{ij}$ is the Kronecker delta operator, and the flow is assumed to be inviscid. This expression thus quantifies the influence of added-mass effects and an unsteady buoyancy force, which comes from the oscillating pressure gradient that drives the oscillating inflow \citep{granlund_airfoil_2014}. We assume that neither the added-mass tensor nor the volume of the body changes as a function of time for a porous disc, and that $U_i(t)$ and $W_i(t)$ are periodic. Since $U_i(t)$ and $W_i(t)$ are periodic, the time averages over a single period of $\frac{dU_j}{dt}$ and $\frac{dW_j}{dt}$ are both zero. It thus follows that the time-averaged force on the body due to these two types of unsteady contributions is also zero, and therefore neither of these unsteady effects creates a theoretical difference between the time-averaged performance of an oscillating turbine and a stationary turbine in an oscillating inflow. We also note that the volume of a 2D actuator disc is effectively zero, and thus the unsteady buoyancy force that differentiates the two scenarios should be negligible.

\section{Experimental Methods}
\label{sec:methods}

\subsection{Experimental Apparatus}

\begin{figure}
\centerline{\includegraphics[width=1.0\textwidth]{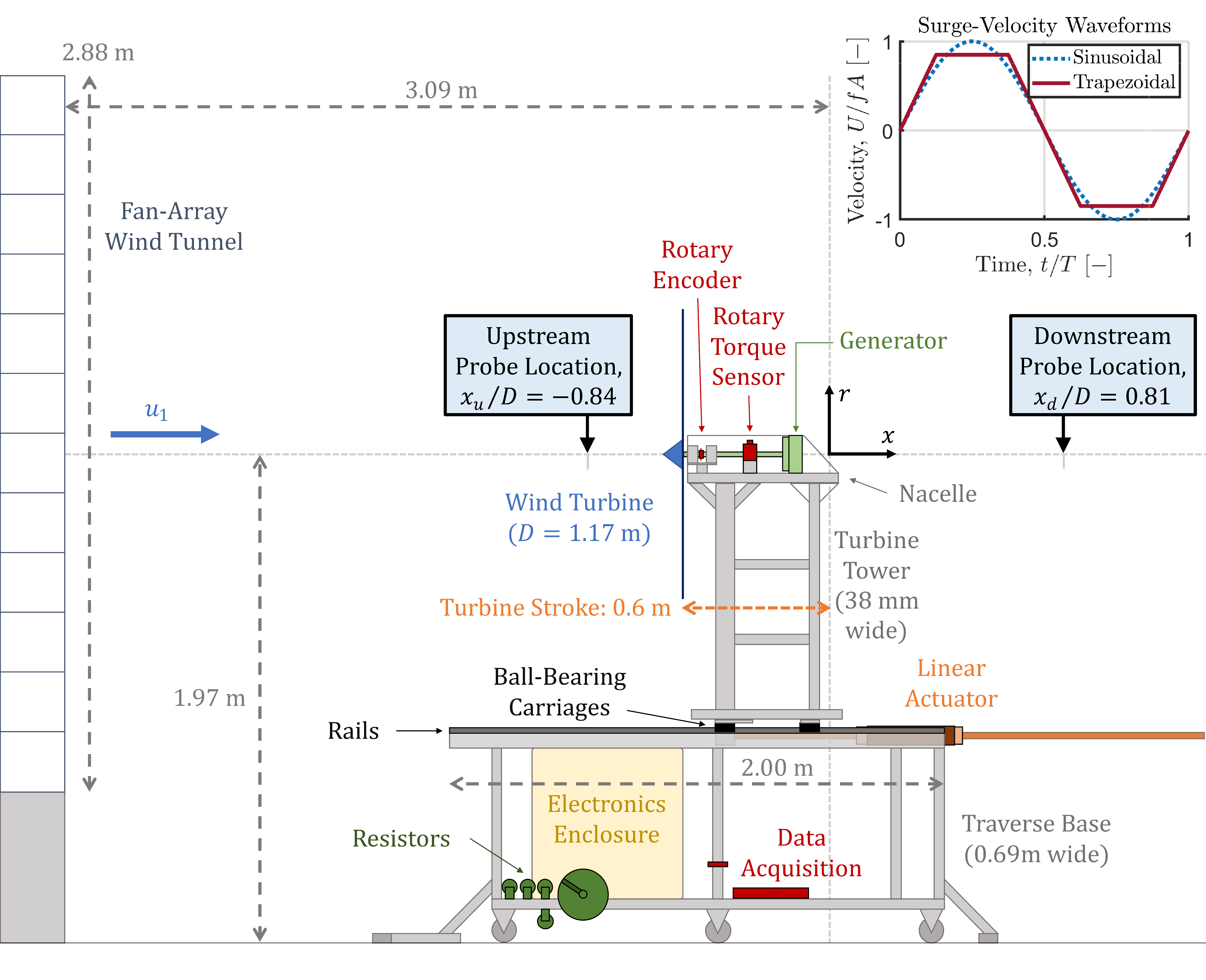}}
\caption{Schematic of the experimental apparatus, including the fan-array wind tunnel (left) and surging turbine (center-right). The turbine is illustrated at its maximum upstream position relative to the origin ($x = -0.6$ m). The inset figure (top right) shows the two types of surge-velocity waveforms used in these experiments.}
  \label{fig:expsetup}
\end{figure}

To characterize the range of conditions over which the ideal-flow model holds, velocity and pressure measurements were conducted in a $2.88\times2.88$ $\rm{m^2}$ open-circuit fan-array wind tunnel at the Caltech Center for Autonomous Systems and Technologies (CAST). A three-bladed, fixed-pitch horizontal-axis wind turbine (Primus Wind Power AIR Silent X) with a rotor diameter of $D = 1.17$ m was mounted on a traverse that translated along 2-m long rails (NSK NH-series) and was actuated by a magnetic piston-type linear actuator (LinMot PS10-70x320U). A diagram of this apparatus is given in Figure \ref{fig:expsetup}. The hub height of the turbine was 1.97 m above the floor of the facility, and the farthest-downstream position of the turbine (defined as $x=0$) was 3.09 m downstream of the fan array. The electrical load on the turbine was provided by 10, 20, and 40-Ohm resistors (TE Connectivity TE1000-series). A rotary torque transducer (FUTEK TRS300) and rotary encoder (US Digital EM2) were used to measure the power produced by the turbine. The estimated blockage of the swept area of the turbine and all support structures, relative to the surface area of the fan array, was 14\%. Further details regarding the dimensions and capabilities of the apparatus may be found in \cite{wei_phase-averaged_2022}.

A constant-temperature hot-wire anemometry system (Dantec MiniCTA 54T42) and differential pressure transducer (MKS Baratron 398-series with Type 270B signal conditioner) were used to measure flow properties at two locations along the turbine centerline, one upstream of the turbine at $x_u = -0.840 D$ and one downstream at $x_d = 0.810 D$. The hot-wire probe was placed approximately on the centerline, while the input line of the pressure transducer was located 3.8 cm to the side. The transducer's reference line was placed in a shielded area outside the flow of the wind tunnel. Data were collected at a sampling rate of 20 kHz and were low-pass filtered using a sixth-order Butterworth filter with a cutoff frequency of 100 Hz. The hot-wire anemometer was calibrated in the wind tunnel against a Pitot probe using the same pressure transducer. Because the facility was exposed to the atmosphere, the temperature and relative humidity were recorded during all experiments to estimate the air density and correct the hot-wire calibration for temperature changes.

\subsection{Experimental Procedure}
\label{sec:procedure}

Experiments were conducted over two nights in March 2022, in which the free-stream velocities in the wind tunnel were $u_1 = 7.79\pm0.10$ and $7.96\pm0.11$ $\rm{ms^{-1}}$, corresponding to an average diameter-based Reynolds number of $Re_D = 6.27\times10^5$. The hot-wire anemometer was calibrated at the beginning and end of each set of experiments. The turbine was operated at three tip-speed ratios, $\lambda_0 = 6.48\pm0.25$, $7.84\pm0.28$, and $8.77\pm0.27$, with corresponding coefficients of power of $C_{p,0} = 0.298\pm0.013\approx C_{p,max}$, $0.248\pm0.012$, and $0.165\pm0.010$. The generator constants for these cases were obtained from the data of \cite{wei_phase-averaged_2022}, as were the turbine and generator moments of inertia ($J=0.0266\pm0.0008$ kg $\rm{m^2}$; $K_2 = 6.96\times10^{-4}$ kg $\rm{m^2}$). The operating parameters of the turbine are summarized in Table \ref{tab:params}. The turbine was actuated in sinusoidal and trapezoidal motions (see inset, Figure \ref{fig:expsetup}) with an amplitude of $A = 0.3$ m ($0.257D$) and periods between $T = 1$ and 6 s, corresponding to nondimensional surge-velocity amplitudes between $u^* \equiv fA/u_1 = 0.039$ and $0.242$. Data were phase-averaged over 100 motion periods. The amplitudes and phases of each quantity of interest were computed from an FFT of the phase-averaged signal. Upstream and downstream flow measurements were collected in separate tests. Additionally, a series of quasi-steady flow measurements were obtained for each tip-speed ratio by placing the turbine at six equally spaced streamwise locations between $x/D = -0.514$ and $0$ and recording measurements over 120 s. To correct against differences in the ambient conditions between measurement sessions and facilitate more direct comparisons, quasi-steady measurements taken at $x/D = 0$ on both sessions were used to scale the measured velocities and pressures from one session to match those from the other session.

\begin{table}
  \begin{center}
  \def~{\hphantom{0}}
  \begingroup
  \renewcommand{\arraystretch}{1.2} 
  \begin{tabular}{l c c c}
        Resistive Load & 10 $\rm{\Omega}$ & 20 $\rm{\Omega}$ & 40 $\rm{\Omega}$ \\
        Case Identifier & 
        \includegraphics[width=10pt]{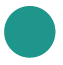} &
        \includegraphics[width=10pt]{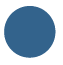} &
        \includegraphics[width=10pt]{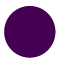} \\\hline
        $\lambda_0$ & $6.48\pm0.25$ & $7.84\pm0.28$ & $8.77\pm0.27$ \\
        $C_{p,0}$ & $0.298\pm0.013$ & $0.248\pm0.012$ & $0.165\pm0.010$ \\\hline
        $K_1$ $\left[\rm{kg\,{m^2}{s^{-1}}}\right]$ & 0.0112 & 0.00649 & 0.00376 \\
        $K_0$ $\left[\rm{Nm}\right]$ & 0.119 & 0.0850 & 0.0676 \\\hline
        $a_0$ (VCT) & 0.252 & 0.267 & 0.221 \\
        $a_0$ (PDT) & 0.286 & 0.301 & 0.250 \\\hline
        $\kappa$ (VCT) & 0.364 & 0.278 & 0.222 \\
        $\kappa$ (PDT) & 0.320 & 0.247 & 0.196 \\
  \end{tabular}
  \endgroup
  \caption{Performance characteristics and model constants for the three loading conditions investigated in this study.}
  \label{tab:params}
  \end{center}
\end{table}

\subsection{Determination of model parameters from steady-flow measurements}

\subsubsection{Power-curve parameterization and model integration}

\begin{figure}
\centerline{\includegraphics[width=0.48\textwidth]{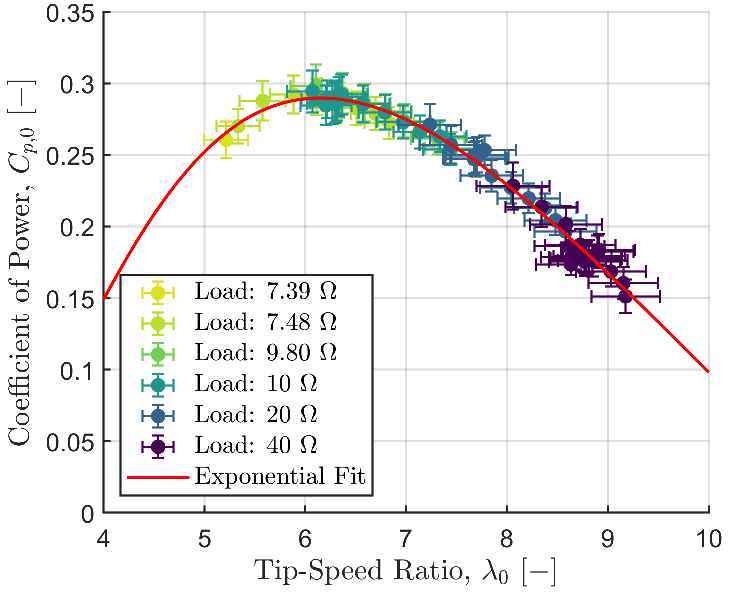}}
\caption{Steady power curve for the turbine used in these experiments, measured over a range of resistive loads and wind speeds. Some of these data points are reproduced from \cite{wei_phase-averaged_2022}. The result of the exponential fit given by Equation \ref{eqn:expCpTSR} is shown as a solid red line.}
  \label{fig:Cp_TSR}
\end{figure}

To compare the analytical modeling framework with the experimental data, it was first necessary to parameterize the steady-flow power curve of the turbine, $C_{p,0}(\lambda)$. The data in the measured power curve of the turbine, shown in Figure \ref{fig:Cp_TSR}, were fitted to a type of exponential function frequently used for wind-turbine modeling:
\begin{equation}
C_p = \left(\frac{c_1}{\lambda+c_2}-c_3\right)\exp{\left(\frac{-c_4}{\lambda+c_2}\right)}.
    \label{eqn:expCpTSR}
\end{equation}
Several coefficients from the general model form given by \cite{heier_grid_2014} were omitted to simplify the model; the four remaining fitted coefficients were $c_1=16.784$, $c_2=-1.510$, $c_3=1.702$, and $c_4=8.764$. This parameterization was employed (as opposed to e.g.\ polynomial fits) to ensure that the slope and concavity at the extremes of the power curve would be captured reliably, since it will be shown in Section \ref{sec:results_pwr} that the performance of the modeling framework is sensitive to these factors.

To obtain time-resolved predictions of the turbine rotation rate, torque, and power, Equation \ref{eqn:model_rotRate} was numerically integrated over ten surge periods using a fourth-order Runge-Kutta scheme. Timesteps were kept no larger than $0.001T$ to maintain numerical stability and accuracy. The steady-flow turbine rotation rate $\omega_0$ was used as the initial condition, and convergence was typically established within a few forcing periods. The model predictions for the amplitude, phase, and time-average of each quantity were computed from the final period in the simulation.

\subsubsection{Steady-flow induction data}

\begin{figure}
\centerline{\includegraphics[width=0.48\textwidth]{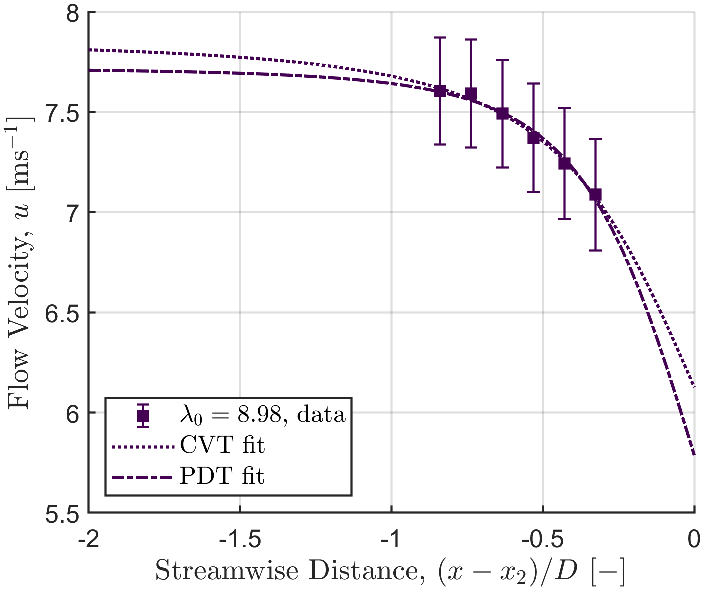}}
\caption{Streamwise-velocity measurements at six different streamwise distances upstream of the turbine for the highest tip-speed ratio tested, compared with two-parameter fits based on the VCT (dotted line) and PDT (dashed-dotted line) induction models. The PDT model shows a slightly more aggressive drop in streamwise velocity close to the turbine and a lower predicted free-stream velocity than the VCT model.}
  \label{fig:compModels}
\end{figure}

\begin{figure}
\begin{subfigure}[t]{0.48\textwidth}
\centering
  \includegraphics[width=\textwidth]{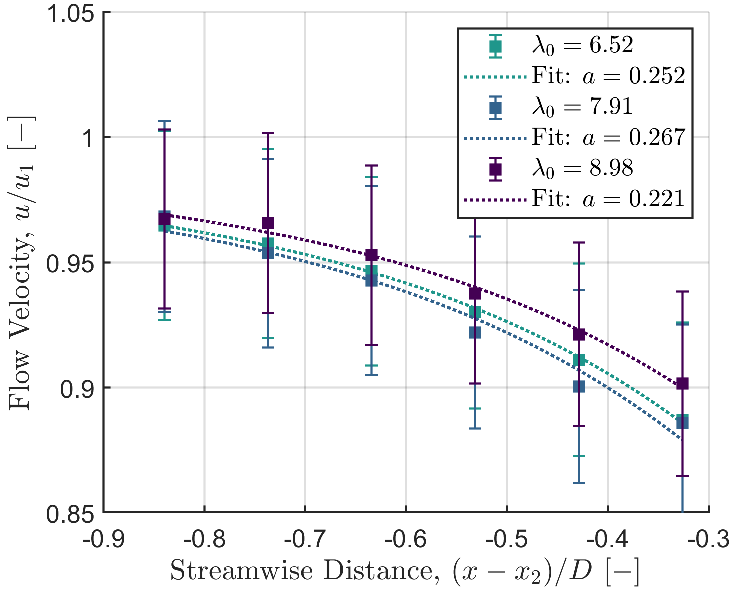}
  \caption{}
\label{fig:u_QS_VCT}
\end{subfigure}
\hfill
\begin{subfigure}[t]{0.48\textwidth}
\centering
  \includegraphics[width=\textwidth]{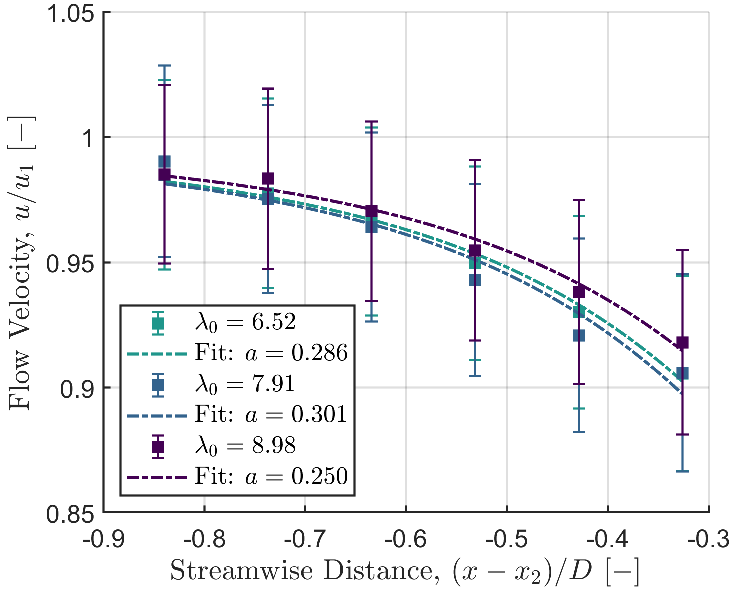}
  \caption{}
\label{fig:u_QS_PDT}
\end{subfigure}
\caption{Streamwise-velocity measurements at six different streamwise distances upstream of the turbine and three tip-speed ratios, compared with fits based on the (a) VCT and (b) PDT induction models. The velocity data are normalized by the free-stream velocities obtained from the two-parameter fits.}
\label{fig:u_QS}
\end{figure}

To estimate the steady induction-factor values of the turbine at the three tip-speed ratios tested in this study, quasi-steady measurements of the streamwise velocity $u(x)$ (described in Section \ref{sec:procedure}) were plotted as a function of streamwise distance $x/D$, and two-parameter fits for the wind speed $u_1$ and centerline induction factor $a_0$ were applied to these data. Dimensional data for the highest tip-speed ratio tested are shown in Figure \ref{fig:compModels} with fits using the VCT (Equation \ref{eqn:VCT}) and PDT (Equation \ref{eqn:PDT}) models. This test case demonstrates the slight differences between the modeling frameworks: for the same data, the PDT model predicts a stronger induction effect (as $x/D \rightarrow 0$) and a slightly lower free-stream velocity ($x/D \leq -2$) than the VCT model. The fit results for all three tip-speed ratios are shown in Figures \ref{fig:u_QS_VCT} (VCT) and \ref{fig:u_QS_PDT} (PDT), where the measured and modeled flow velocities are normalized by the wind speeds identified by each two-parameter fit. These steady-flow tests demonstrate that, within the range of streamwise distances tested, the agreement of both models with the trends observed in the data is reasonably good.

The centerline induction-factor values shown in Figure \ref{fig:u_QS} were compared with the rotor-averaged values estimated using Equation \ref{eqn:CpSurge} and a $C_p$ calculated from the average of the measured turbine torque and rotation rate over all six streamwise locations. The ratio between these estimates of $\langle a_0 \rangle$ and the fitted values of $a_0$ gave an empirical estimate for the correction factor $\kappa$ for each tip-speed ratio. The estimated values of $a_0$ and $\kappa$ are given in Table \ref{tab:params} for the two induction-model frameworks. For comparison, an analytical correction factor can be calculated by integrating the self-similar solution of \cite{troldborg_simple_2017}, which yields $\kappa=0.649$. This is larger than the empirically estimated values of $\kappa$. However, the ratio of the turbine hub diameter to the turbine diameter for the simulations used to calibrate the self-similar solution was 2.4\%, whereas for the turbine used in this study the ratio was around 12\%. It is therefore expected that the flow would decelerate more strongly along the centerline of this particular turbine, due to the increased blockage effect from the larger hub and nacelle, thus lowering the ratio between the rotor-averaged and centerline induction factors.

\subsubsection{Wind-speed and pressure corrections}

Since changes in the turbine tip-speed ratio correspond to changes in the thrust force on the turbine, the operation of the turbine in the open test section of the wind tunnel created a blockage effect that influenced the wind speed in the open test section -- an effect that is well-documented in the literature \cite[e.g.][]{eltayesh_effect_2019}. To correct against this additional source of error, a wind-speed correction was computed for each unsteady test case by comparing the mean of the streamwise-velocity measurements with the wind speeds measured by the hot-wire anemometer during calibration (where the turbine was present in the tunnel but was not rotating). The average of these fitted wind speeds across all of the unsteady and quasi-steady tests for each tip-speed ratio was then used as an adjusted wind speed for normalizing the recorded velocity and pressure data. 

Further corrections were implemented to reduce the influence of sources of uncertainty in the pressure data. Since the absolute pressure at the reference of the pressure transducer was unknown, the mean value of this pressure prediction was scaled to be equal to that of the data. Additionally, due to the long length of the tubes that connected the pressure transducer to the measurement location, a first-order low-pass filter with a cutoff frequency of 2.48 Hz was inferred from the phase of the measured pressure data relative to the velocity signal. This filter was then applied to the calculated model predictions for pressure. Remaining discrepancies between the measured and modeled pressure signals could be attributed to the true filtering effect of the tubes being of higher order than the first-order filter model \citep{bergh_theoretical_1965}.

\section{Experimental Results}
\label{sec:results}

In this section, the predictions of the nonlinear dynamical model derived in Section \ref{sec:model_pwr} are compared with experimental measurements of time-averaged and fluctuating quantities. The induction-factor estimates collected from the nonlinear dynamical model are then used to predict the streamwise velocity and pressure upstream of the surging turbine. Predictions based on vortex-cylinder theory (VCT) and porous-disc theory (PDT) are compared with flow measurements to demonstrate that the proposed modeling framework is able to reproduce trends observed in the measured response of the system.

\subsection{Power generation}
\label{sec:results_pwr}

The predictions of the nonlinear ordinary differential equation given in Equation \ref{eqn:model_rotRate} are compared with rotation-rate, torque, and power data collected from the surging-turbine experimental apparatus. For this purpose, we reuse the experimental torque and power measurements of \cite{wei_phase-averaged_2022} and plot the results of the present modeling framework against these data. This dataset contained measurements at three lower tip-speed ratios, in addition to those investigated in the present study, and measurements were conducted over a wider range of surge-velocity amplitudes and frequencies.

\begin{figure}
\begin{subfigure}[t]{0.48\textwidth}
\centering
  \includegraphics[width=\textwidth]{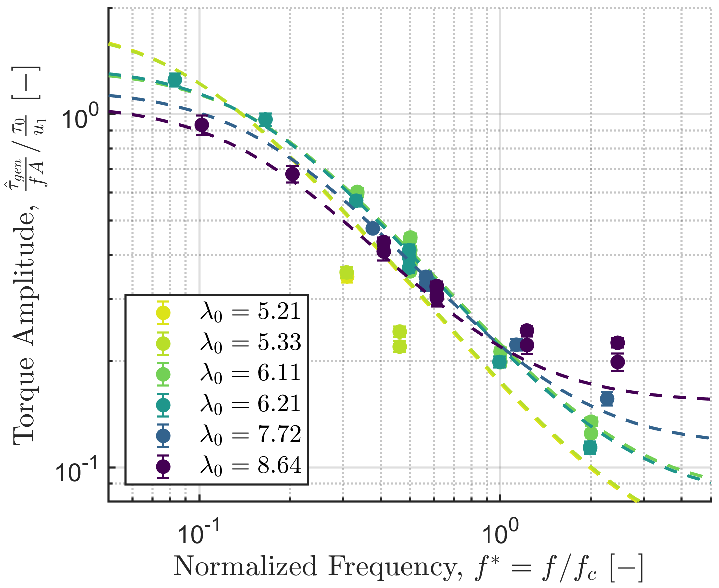}
  \caption{}
\label{fig:Tau_amp}
\end{subfigure}
\hfill
\begin{subfigure}[t]{0.48\textwidth}
\centering
  \includegraphics[width=\textwidth]{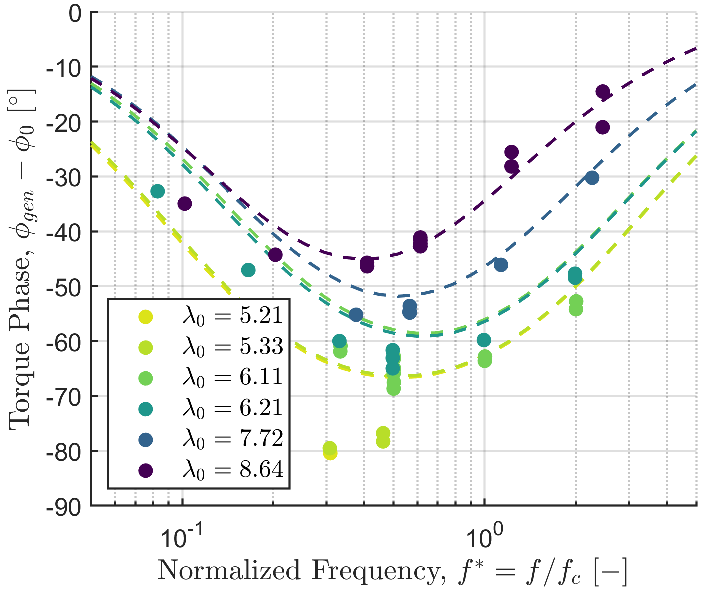}
  \caption{}
\label{fig:Tau_phase}
\end{subfigure}
\caption{Generator-torque (a) amplitude and (b) phase for a series of sinusoidal surge-velocity waveforms, plotted against normalized frequency and compared with model predictions (dashed lines). Data are reproduced from \cite{wei_phase-averaged_2022}.}
\label{fig:Tau}
\end{figure}

Figure \ref{fig:Tau} shows the measured torque amplitude and phase from these data, plotted against surge frequency. The torque amplitude was scaled by the surge-velocity amplitude $u^*=fA$ in the manner of a transfer-function gain, and this was nondimensionalized by the steady reference torque $\tau_{gen,0}$ and the free-stream velocity. The frequency was nondimensionalized by a characteristic frequency $f_c$ derived from the linear model of \cite{wei_phase-averaged_2022}. The trends in the data are relatively well-captured by the model predictions, and the nonlinear model shows some improvement over the linear model at the lowest tip-speed ratios tested \cite[cf.][Figures 7a and 9a]{wei_phase-averaged_2022}. This suggests that the nonlinear model derived in this work is an effective generalization of the linearization developed and validated in the preceding study.

\begin{figure}
\begin{subfigure}[t]{0.48\textwidth}
\centering
  \includegraphics[width=\textwidth]{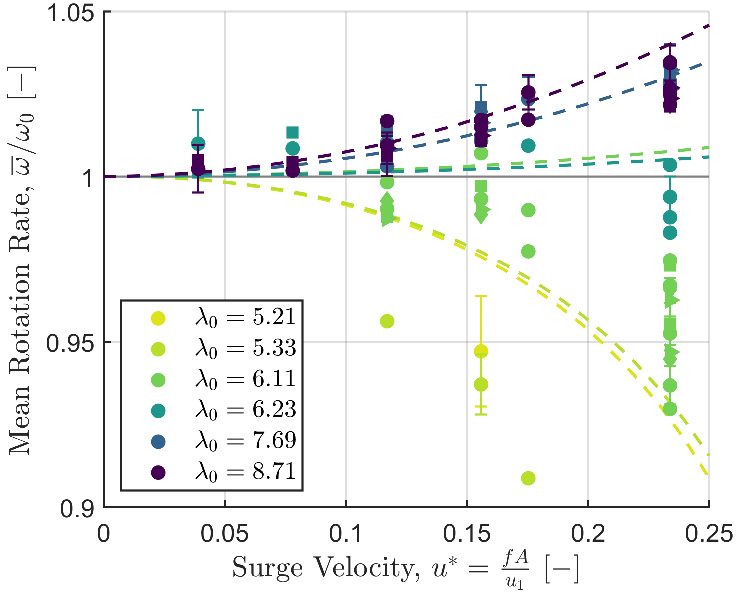}
  \caption{}
\label{fig:omega_vs_u}
\end{subfigure}
\hfill
\begin{subfigure}[t]{0.48\textwidth}
\centering
  \includegraphics[width=\textwidth]{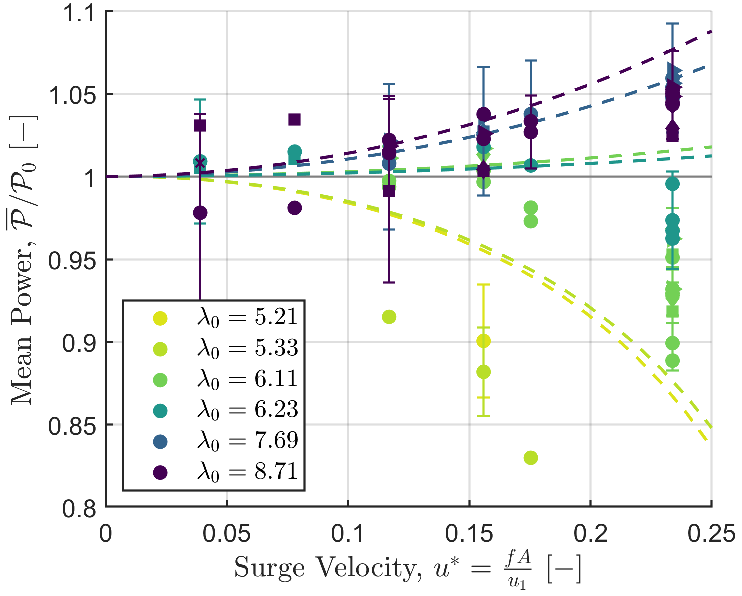}
  \caption{}
\label{fig:Pwr_vs_U}
\end{subfigure}
\caption{Time-averaged (a) rotation rate and (b) power, plotted against surge-velocity amplitude. Data are reproduced from \cite{wei_phase-averaged_2022}; model predictions derived from Equation \ref{eqn:model_rotRate} are plotted as dashed lines and colored by tip-speed ratio. Circles represent sinusoidal surge-velocity waveforms and diamonds and other markers represent trapezoidal waveforms. Error bars are plotted on every sixth point for the sake of clarity.}
\label{fig:timeAvg}
\end{figure}

The benefit of the nonlinear model is evident when predictions for time-averaged quantities of interest are required. According to the preceding linearized model, the time-averaged rotation rate and power of the turbine will not deviate from their corresponding steady-flow quantities. Departures from this ansatz are clearly evident in Figure \ref{fig:timeAvg}, as the mean rotation rate and power decrease as a function of increasing surge-velocity amplitude for low tip-speed ratios, and increase for high tip-speed ratios. The predictions of the nonlinear model, however, are able to follow these trends. The largest enhancements in time-averaged power over the steady case are predicted at the highest tip-speed ratios, while the greatest decreases in time-averaged power occur at the lowest tip-speed ratios. The nonlinear model still overpredicts these power decreases; it is likely that these additional losses are a result of the inception of flow separation on the turbine blades. The turbine stalled and stopped spinning when forced to operate at or below $\lambda_0\approx 5$, and so the decrease in power as a function of decreasing tip-speed ratio was in reality much sharper than that suggested by the fit to the turbine power curve in Figure \ref{fig:Cp_TSR}. The lack of a parameterization for these dynamics in the current modeling framework is thus probably responsible for the lack of quantitative agreement at lower tip-speed ratios.

The discussion of flow separation and the extent to which stall is captured in the turbine power curve emphasizes the key point that the critical nonlinearity in Equation \ref{eqn:model_rotRate} is the functional form of the power curve itself. In other words, the present quasi-steady modeling framework hinges on a reliable characterization of the turbine in steady flow conditions. This is a considerable advantage of the modeling approach, since it precludes the need for unsteady calibration and can thus be applied directly to the design of turbines in unsteady flow conditions when only steady-flow data are available. It also implies, however, that particular attention must be paid to the parameterization of the steady-flow power curve of the turbine. This dependence and its implications will be discussed in Section \ref{sec:enhancements}.

\subsection{Upstream flow properties}
\label{sec:results_flow}

\begin{figure}
\begin{subfigure}[t]{0.48\textwidth}
\centering
  \includegraphics[width=\textwidth]{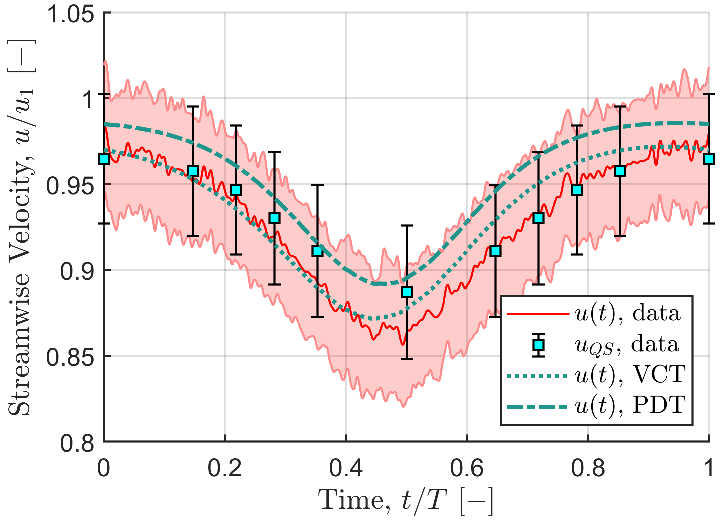}
  \caption{}
\label{fig:uHW_10_sine}
\end{subfigure}
\hfill
\begin{subfigure}[t]{0.48\textwidth}
\centering
  \includegraphics[width=\textwidth]{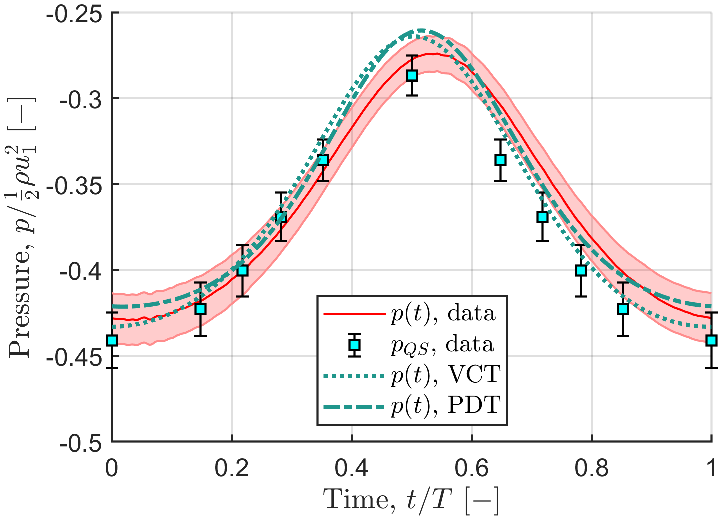}
  \caption{}
\label{fig:p_10_sine}
\end{subfigure}
\begin{subfigure}[t]{0.48\textwidth}
\centering
  \includegraphics[width=\textwidth]{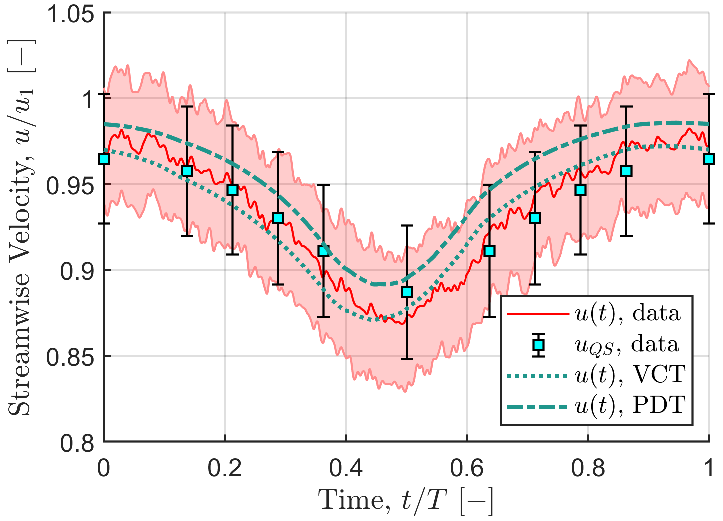}
  \caption{}
\label{fig:uHW_10_trap}
\end{subfigure}
\hfill
\begin{subfigure}[t]{0.48\textwidth}
\centering
  \includegraphics[width=\textwidth]{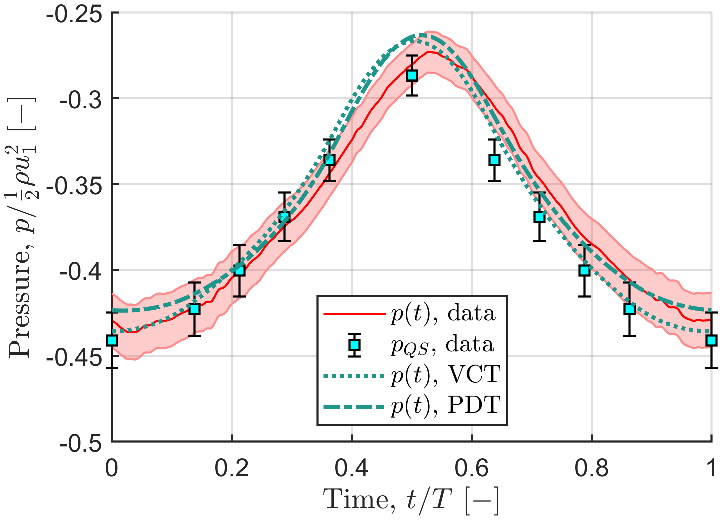}
  \caption{}
\label{fig:p_10_trap}
\end{subfigure}
\begin{subfigure}[t]{0.48\textwidth}
\centering
  \includegraphics[width=\textwidth]{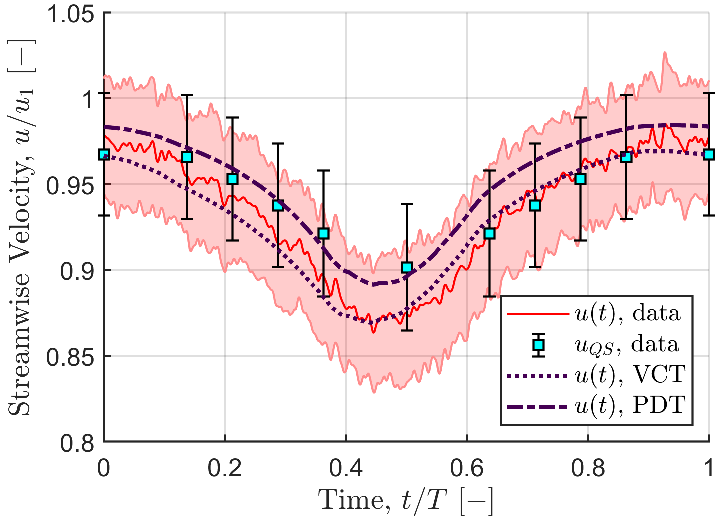}
  \caption{}
\label{fig:uHW_40_trap}
\end{subfigure}
\hfill
\begin{subfigure}[t]{0.48\textwidth}
\centering
  \includegraphics[width=\textwidth]{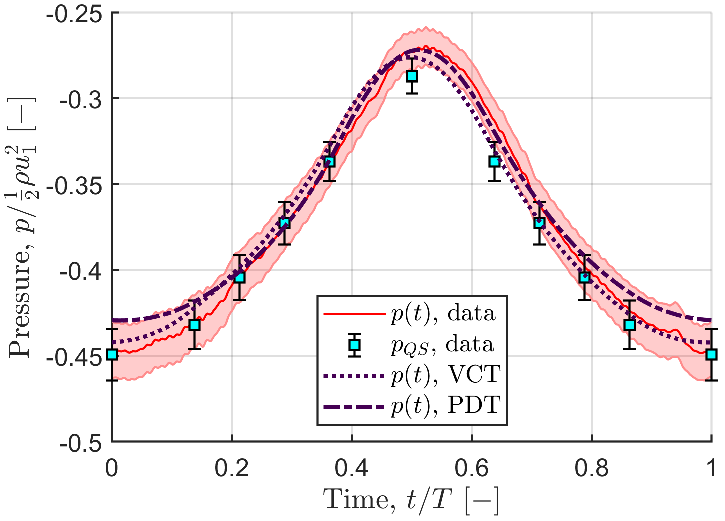}
  \caption{}
\label{fig:p_40_trap}
\end{subfigure}
\caption{Phase-averaged velocity and pressure profiles for (a,b) a sinusoidal surge-velocity waveform with $\lambda_0=6.48$, (c,d) a trapezoidal waveform with $\lambda_0=6.48$, and (e,f) a trapezoidal waveform with $\lambda_0=8.77$. All surge-velocity waveforms had $u^* = 0.242$. The solid red lines represent unsteady measurements, the blue squares represent quasi-steady measurements, the dotted lines show the VCT model predictions, and the dashed-dotted lines show the PDT model predictions.}
\label{fig:phaseAveraged}
\end{figure}

We now investigate the extension of the nonlinear model for the turbine dynamics to the flow properties in the upstream induction region of the turbine. The unsteady and quasi-steady data from three selected experimental cases, all measured at $x=x_u$, are shown in Figure \ref{fig:phaseAveraged}. The streamwise-velocity signals showed a phase lead and increased amplitude with respect to the quasi-steady measurements. Also shown in these figures are the VCT and PDT model predictions, which align well with the shape of the phase-averaged data and anticipate the increased amplitudes and phase leads as well. For all of these cases, both models show good agreement with the velocity and pressure data, as well as with each other.


\begin{figure}
\begin{subfigure}[t]{0.48\textwidth}
\centering
  \includegraphics[width=\textwidth]{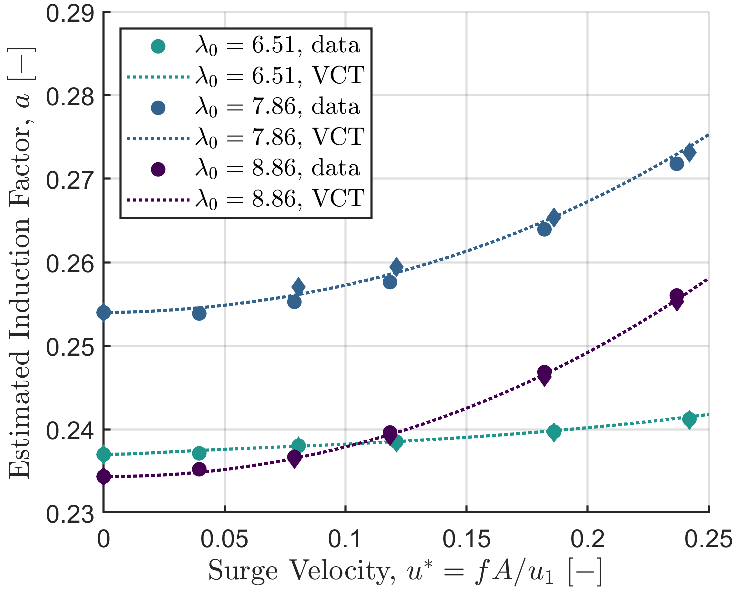}
  \caption{}
\label{fig:a_vs_u_VCT}
\end{subfigure}
\hfill
\begin{subfigure}[t]{0.48\textwidth}
\centering
  \includegraphics[width=\textwidth]{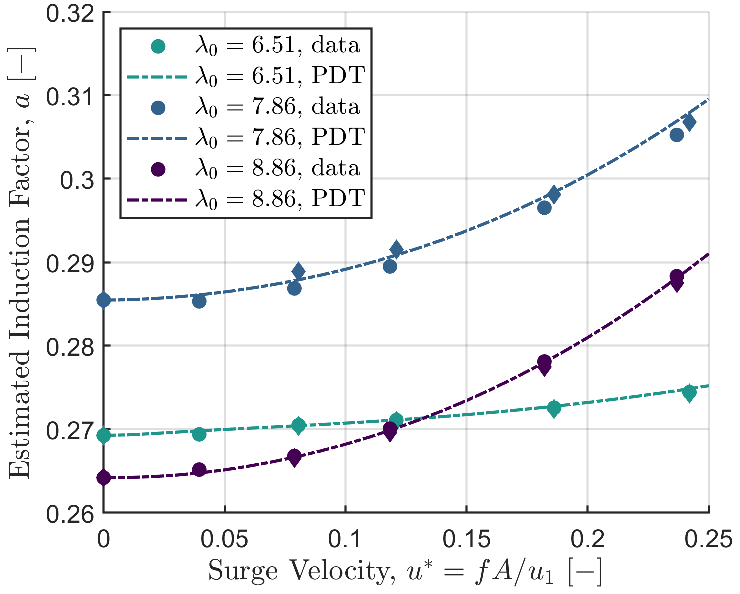}
  \caption{}
\label{fig:a_vs_u_PDT}
\end{subfigure}
\caption{Time-averaged induction-factor estimates from the (a) VCT and (b) PDT models. For these and all following figures, circles indicate sinusoidal-waveform data and diamonds denote trapezoidal-waveform data. Points are estimated from measured power data, while lines are estimated from the nonlinear dynamical model of the turbine.}
\label{fig:a_vs_u}
\end{figure}

The differences between the induction predictions of the models is reflected in the induction factors estimated from the power data. The time-averaged induction factors estimated using the VCT and PDT models and correction factors are shown in Figures \ref{fig:a_vs_u_VCT} and \ref{fig:a_vs_u_PDT}, respectively. The induction factors all increase with increasing surge-velocity amplitude, but at different rates depending on tip-speed ratio. The trends are consistent between the two models; the main difference is that the estimated induction factors from the PDT model are slightly higher, in accordance with the model's sharper induction profile (noted previously in Figure \ref{fig:compModels}). 

The increases in time-averaged induction factor with surge-velocity amplitude are tied to increases in the thrust force exerted by the turbine on the incoming flow. According to 1D momentum theory, the thrust coefficient of the turbine is given as 
\begin{equation}
    C_t \equiv \frac{F_t}{\frac{1}{2}\rho \pi R^2 u_\infty^2}= 4\aRA\left(1-\aRA\right),
    \label{eqn:CT}
\end{equation}
which, for the relatively low values of $\aRA$ considered in this study, increases with increasing $\aRA$. The induction-factor estimates in Figure \ref{fig:a_vs_u} therefore suggest that the thrust force on the turbine increases with increasing surge-velocity amplitude.

This aligns with a simplified analysis of the thrust force of a turbine in an oscillating inflow. If $C_t$ is held constant and the dimensional thrust is integrated over a sinusoidal surge-velocity waveform, as done for $C_p$ in Equation \ref{eqn:powerWenJohlas}, the time-averaged thrust enhancement is
\begin{equation}
\frac{\overline{F_t}}{F_{t,0}} = 1 + \frac{1}{2}{u^*}^2,
\label{eqn:thrustUnsteadyAvg}
\end{equation}
which is also an increasing function with $u^*$. Although direct thrust measurements were not possible in this study due to the large inertial forces associated with the turbine motions, the induction-factor estimates and constant-$C_t$ analysis both suggest that unsteady surge motions increase the time-averaged thrust loading on the turbine rotor. In accordance with the trends observed in the time-averaged power data in Figure \ref{fig:Pwr_vs_U}, the time-averaged thrust may decrease with increasing surge-velocity amplitudes for tip-speed ratios lower than those investigated here.

\begin{figure}
\begin{subfigure}[t]{0.48\textwidth}
\centering
  \includegraphics[width=\textwidth]{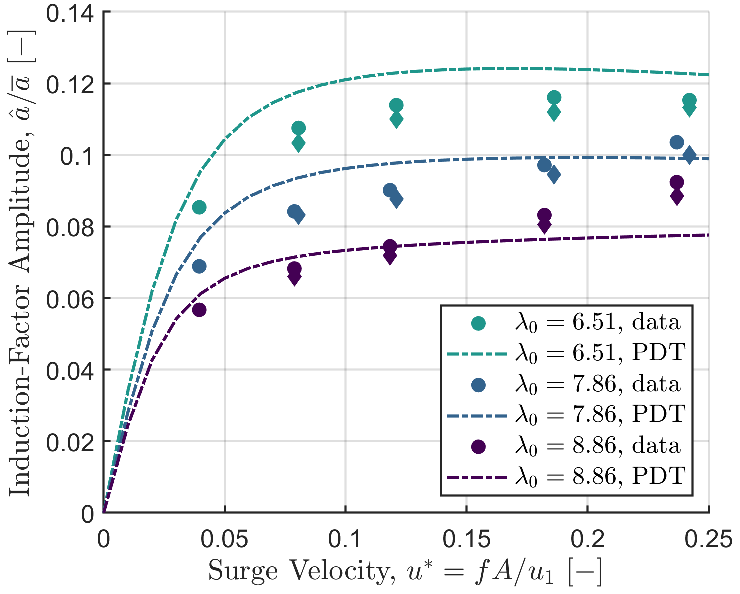}
  \caption{}
\label{fig:a_amp}
\end{subfigure}
\hfill
\begin{subfigure}[t]{0.48\textwidth}
\centering
  \includegraphics[width=\textwidth]{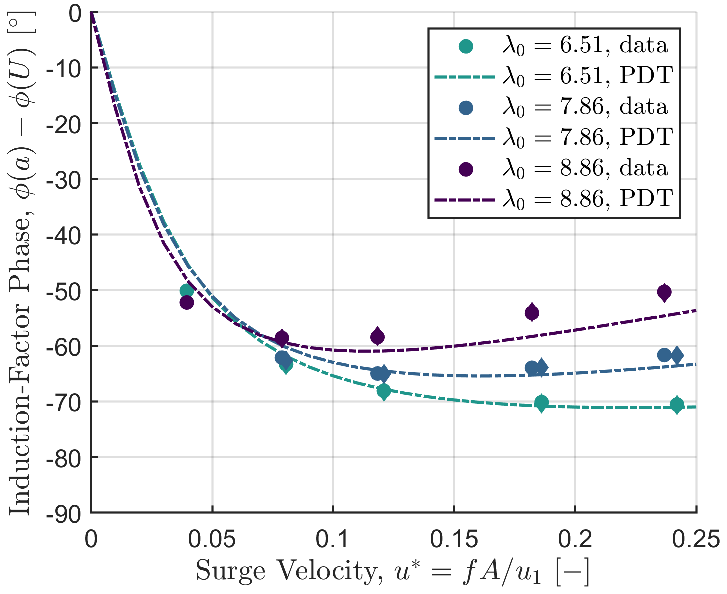}
  \caption{}
\label{fig:a_phase}
\end{subfigure}
\caption{(a) Amplitude and (b) phase of the estimated induction factors using the PDT model, plotted against surge-velocity amplitude. Model predictions are given as dashed-dotted lines.}
\label{fig:a_TF}
\end{figure}

\begin{figure}
\begin{subfigure}[t]{0.48\textwidth}
\centering
  \includegraphics[width=\textwidth]{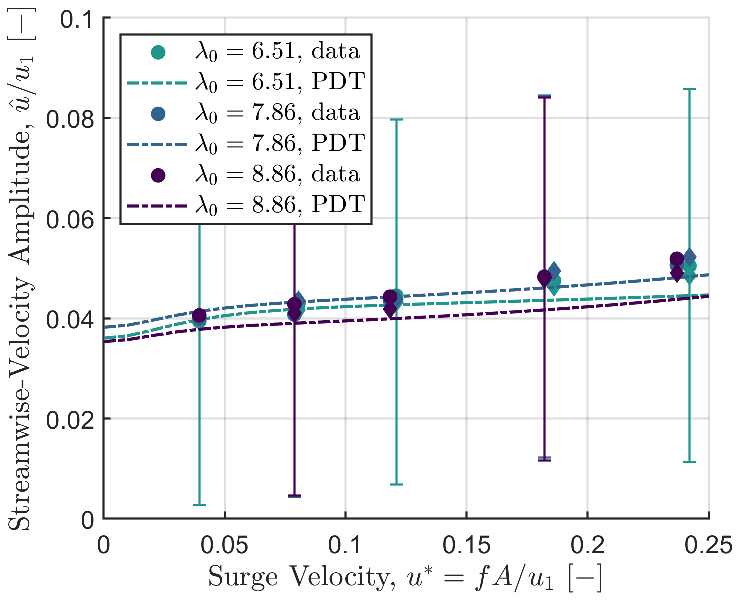}
  \caption{}
\label{fig:uHW_amp}
\end{subfigure}
\hfill
\begin{subfigure}[t]{0.48\textwidth}
\centering
  \includegraphics[width=\textwidth]{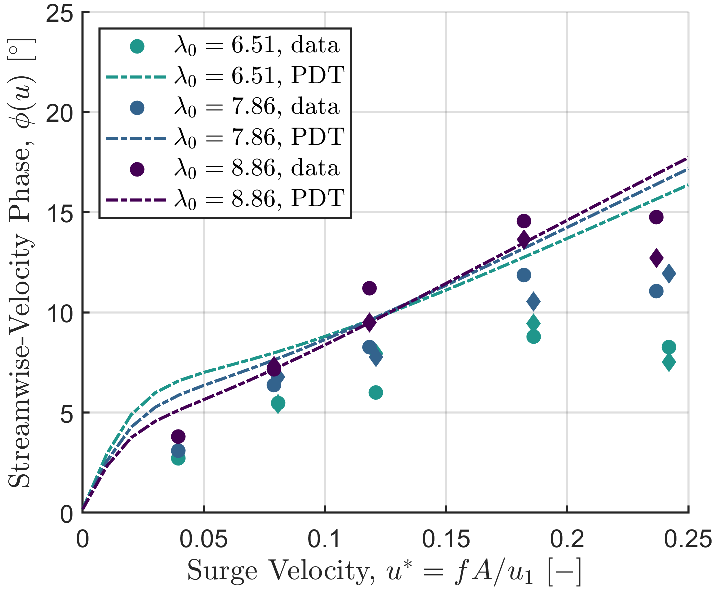}
  \caption{}
\label{fig:uHW_phase}
\end{subfigure}
\caption{(a) Amplitude and (b) phase of the measured flow velocity at $x=x_u$, plotted against surge-velocity amplitude. PDT model predictions are given as dashed-dotted lines. Error bars are plotted on every fourth point for clarity.}
\label{fig:uHW_TF}
\end{figure}

\begin{figure}
\begin{subfigure}[t]{0.48\textwidth}
\centering
  \includegraphics[width=\textwidth]{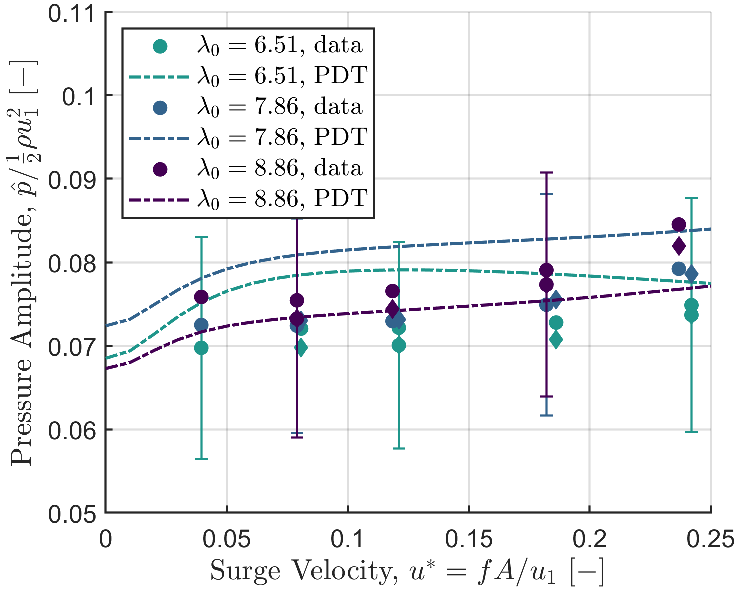}
  \caption{}
\label{fig:p_amp}
\end{subfigure}
\hfill
\begin{subfigure}[t]{0.48\textwidth}
\centering
  \includegraphics[width=\textwidth]{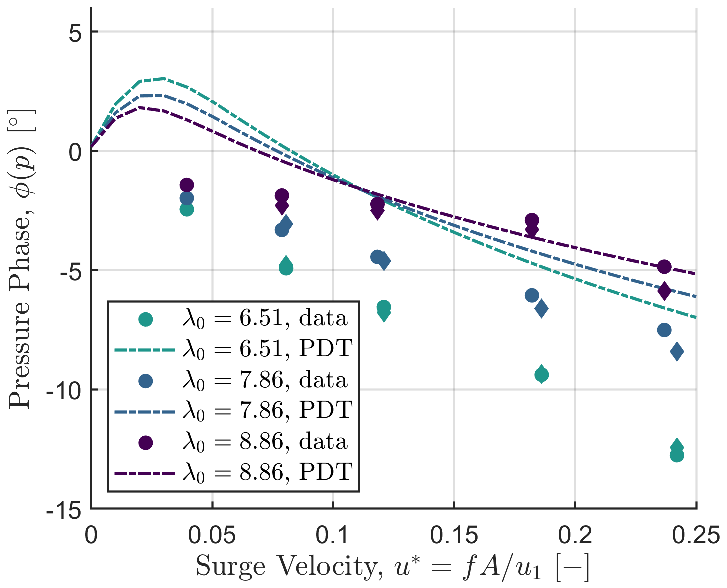}
  \caption{}
\label{fig:p_phase}
\end{subfigure}
\caption{(a) Amplitude and (b) phase of the measured relative pressure at $x=x_u$, plotted against surge-velocity amplitude. PDT model predictions are given as dashed-dotted lines. Error bars are plotted on every fourth point.}
\label{fig:p_TF}
\end{figure}

The case studies in Figure \ref{fig:phaseAveraged} and induction-factor estimates in Figure \ref{fig:a_vs_u} suggest that the VCT and PDT models produce very similar results. For the sake of clarity, Figures \ref{fig:a_TF} through \ref{fig:p_TF} will only show PDT predictions. The VCT predictions are qualitatively similar, and are provided for completeness in Appendix \ref{sec:app_VCT}.

In Figure \ref{fig:a_TF}, the amplitude and phase (relative to the surge-velocity waveform) of the estimated induction factors are presented as a function of surge-velocity amplitude. The induction-factor amplitudes and phase offsets decrease in magnitude with increasing tip-speed ratio; the highest amplitudes observed in the measured data do not exceed 12\% of the time-averaged values. Good agreement between the estimates from the measured data and the estimates from the nonlinear dynamical model is observed.

Good agreement between the model predictions and measured data is also observed in the streamwise-velocity data, shown in terms of amplitude and phase (relative to the quasi-steady measurements) in Figure \ref{fig:uHW_TF}. As noted previously, the surge motions slightly increase the amplitude of the velocity oscillations above the quasi-steady case, and a slight phase lead accrues with increasing surge-velocity amplitude. There is some quantitative mismatch between the phase measurements and model predictions, including a plateau in the phase data at the highest surge-velocity amplitude that is not reflected by the model. Given the large experimental uncertainties in the streamwise-velocity measurements, this is not surprising. However, the slopes of the model-prediction lines still align with the slopes of the measured data (except for the highest surge-velocity amplitude tested), which differ slightly across tip-speed ratios. This suggests that the modeling framework is able to capture some of the more subtle differences in flow properties in the upstream induction region as the tip-speed ratio of the turbine changes.

Similar agreement in trend is observed in the pressure data, shown in amplitude and phase (relative to the quasi-steady measurements) in Figure \ref{fig:p_TF}. The slight increase in pressure amplitude as a function of surge-velocity amplitude is generally reflected in the model predictions, and as in the streamwise-velocity phase data, the model predictions of the pressure phase follow similar slopes to those evident in the data. Again, some quantitative differences are apparent in the plots, but given the measurement uncertainties and the relatively small magnitude of the signals being quantified and predicted, the qualitative agreement between the model predictions and measured data suggests that the modeling framework is parameterizing the salient dynamics of the system.

\begin{figure}
\begin{subfigure}[t]{0.48\textwidth}
\centering
  \includegraphics[width=\textwidth]{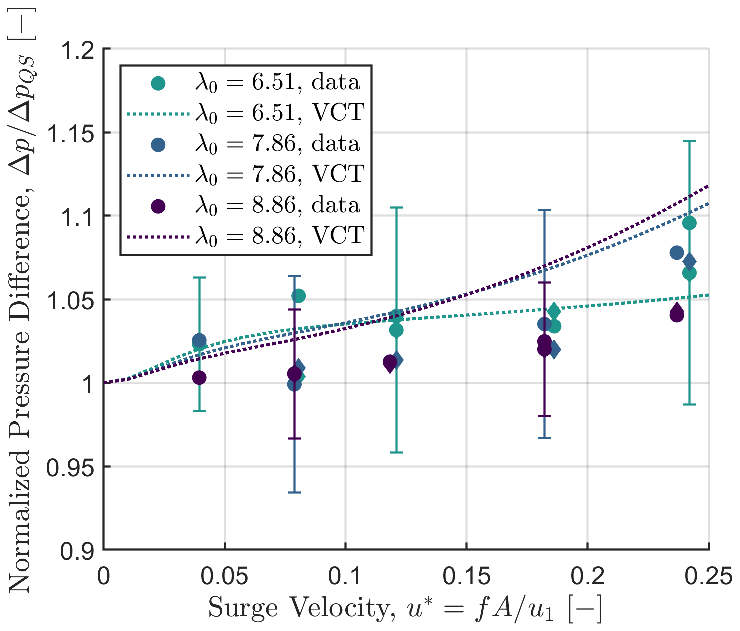}
  \caption{}
\label{fig:dP_VCT}
\end{subfigure}
\hfill
\begin{subfigure}[t]{0.48\textwidth}
\centering
  \includegraphics[width=\textwidth]{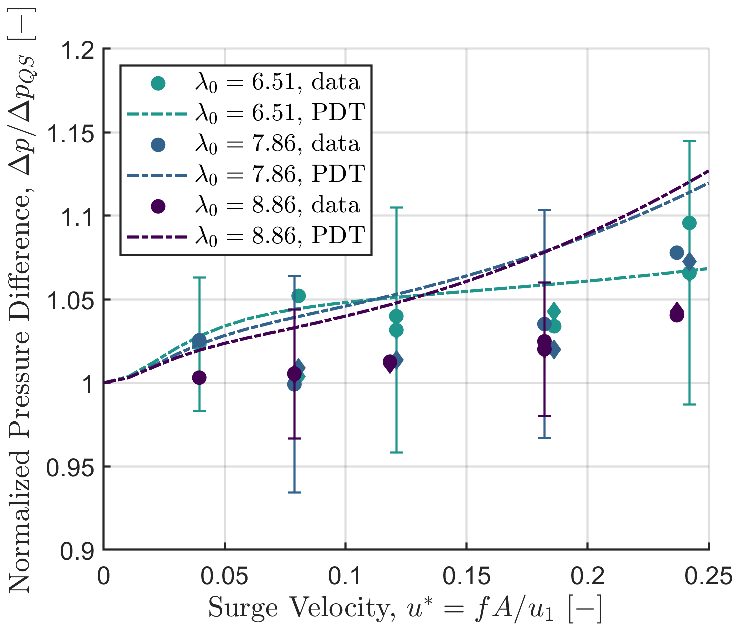}
  \caption{}
\label{fig:dP_PDT}
\end{subfigure}
\caption{Pressure difference across the turbine, $\Delta p = p(x=x_u)-p(x=x_d)$, normalized by the quasi-steady pressure difference and plotted against surge-velocity amplitude. Model predictions are given as dotted lines for the VCT model in (a) and as dashed-dotted lines for the PDT model in (b). Error bars are shown on every fourth point.}
\label{fig:dP}
\end{figure}

Finally, the time-averaged difference between the upstream and downstream pressure measurements at $x=x_u$ and $x=x_d$, respectively, is shown as a function of surge-velocity amplitude in Figure \ref{fig:dP} for both the VCT and PDT models. The corresponding model predictions are calculated by evaluating the model at $x=x_u$ and $x=-x_d$, and assuming that the pressure exhibits odd symmetry about the turbine rotor plane so that $p(x=-x_d) = -p(x=x_d)$, as is the case for 1D momentum theory. For a real turbine with a low-pressure wake region, this is likely not a tenable assumption. However, the model predictions still follow the trends in the data within measurement uncertainty. The measured pressure difference can be understood as an analogue to the thrust force on the turbine, and as previously implied by the time-averaged induction-factor estimates, it increases with increasing surge-velocity amplitude. As well as further demonstrating the predictive capabilities of the modeling framework, these data suggest that the model could also be extended to serve as an initial condition for wake models of the surging turbine, which will depend on the thrust force and pressure drop across the rotor.

In summary, both induction-model frameworks are able to reproduce the trends observed in the centerline flow measurements recorded upstream of the turbine, and the modeling approach may have some bearing on the near-wake region downstream of the turbine as well. This observed agreement implies that the proposed modeling framework is capturing the dominant dynamics of the surging-turbine system, in spite of only incorporating empirical data from the turbine power curve and the induction-factor correction parameter $\kappa$ -- both of which can be obtained from steady-flow theories, simulations, or experiments. While the nonlinear dynamical model accounts for unsteady rotation accelerations of the turbine, the aerodynamics of the turbine itself are treated in a quasi-steady manner, as a function of the power curve and the induction models. The fact that this quasi-steady first-principles model still manages to align with data collected from a real surging turbine might suggest that unsteady flow physics are not strongly present in this system. However, the uncertainties in these experiments preclude us from ruling out this possibility, and we will thus present a theoretical analysis of the effects of unsteady flows in Section \ref{sec:unsteady}.

\section{Further Theoretical Considerations}
\label{sec:discussion}

The experimental results presented in Section \ref{sec:results} establish the predictive capabilities of the modeling framework outlined in Section \ref{sec:model}. In this section, we address remaining considerations regarding the sensitivity of the unsteady turbine performance to its steady-flow aerodynamics and the influence of unsteady flow physics on the system. These analyses highlight potential design strategies for maximizing unsteady power enhancements with wind-energy systems in dynamic-inflow conditions.

\subsection{Time-averaged power enhancements}
\label{sec:enhancements}

The results presented in Section \ref{sec:results_pwr} suggested that the amount of enhancement in the time-averaged power extraction of a surging turbine relative to the steady case depends on the characteristics of the power curve of the turbine. To explore this connection further, a parametric survey using simulations of the turbine model from Equation \ref{eqn:model_rotRate} with simplified power curves was instantiated. For four known equilibrium points along the power curve, corresponding to four of the tip-speed ratios tested in experiments, local quadratic power curves were constructed with the form
\begin{equation}
    C_p(\lambda) = C_p(\lambda_0) + \frac{dC_p}{d\lambda}\vert_{\lambda_0} (\lambda-\lambda_0) + \frac{1}{2}\frac{d^2C_p}{d\lambda^2}(\lambda-\lambda_0)^2,
\end{equation}
where the local slope $\frac{dC_p}{d\lambda}\vert_{\lambda_0}$ was taken from the fitted power curve in Equation \ref{eqn:expCpTSR} and the concavity $\frac{d^2C_p}{d\lambda^2}$ was varied between -0.1 and 0.02. The four tip-speed ratios used in this exploration were $\lambda_0=5.47$, 6.33, 7.67, and 8.69, and the local slope of the second tip-speed ratio was approximated as zero. This parameterization created a set of power curves for each equilibrium point with identical local slope but different concavities, and the effect of changing local slope could be ascertained by comparing the simulation results across tip-speed ratios. A fixed surge-velocity amplitude of $u^*=0.24$ was used for the simulations; other relevant parameters were $u_1 = 8$ $\rm{ms^{-1}}$, $\rho=1.19$ $\rm{kgm^{-3}}$, and $\Delta t = 0.001$ s. As before, a fourth-order Runge-Kutta scheme was used to integrate the model forward in time over ten surge-oscillation periods. A strict convergence metric required the difference between successive periods to decrease monotonically; test cases that failed this criterion were deemed unstable and were not plotted.

\begin{figure}
\centerline{\includegraphics[width=0.48\textwidth]{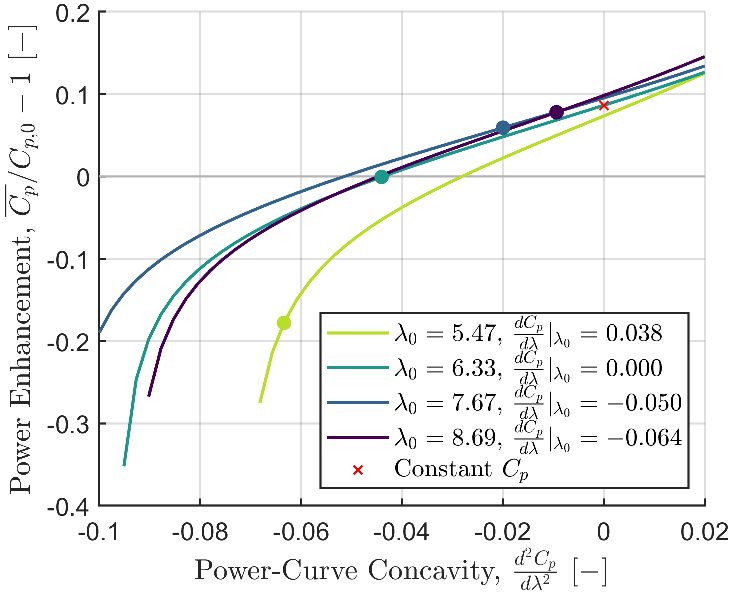}}
\caption{Fractional enhancements in the time-averaged coefficient of power, plotted against the concavity of the power curve (approximated as a quadratic function with fixed local slope and variable concavity), for four tip-speed ratios. A sinusoidal surge-velocity waveform with $u^*=0.24$ was used for these simulations. The red $\times$ shows the constant-$C_p$ solution given in Equation \ref{eqn:powerWenJohlas}, and the solid circles denote the approximate concavity of the actual power curve at each reference tip-speed ratio.}
  \label{fig:concavity}
\end{figure}

The results of these simulations are shown in Figure \ref{fig:concavity}, where the relative power enhancement $\overline{C_p}/C_{p,0} - 1$ is plotted as a function of power-curve concavity for the four selected tip-speed ratios. Filled circles show the concavity of the full power curve from Equation \ref{eqn:expCpTSR}, corresponding approximately to the power gains and losses observed in Figure \ref{fig:Pwr_vs_U} at $u^*\approx 0.24$. At zero slope and zero concavity, the system reduces to the quasi-steady prediction derived by \cite{wen_influences_2017} and \cite{johlas_floating_2021} (Equation \ref{eqn:powerWenJohlas}, shown as a red $\times$ in the figure). Where the concavity is zero, differences are still evident across local slopes: the two highest tip-speed ratios have negative local slopes and larger power enhancements relative to the constant-$C_p$ case. Conversely, the lowest tip-speed ratio (which has a positive local slope) exhibits the lowest power enhancement. As concavity decreases from zero, these power gains decrease and eventually become losses, finally becoming unstable below some critical concavity. Positive concavities, by contrast, show increasing power enhancements. 

The influence of power-curve concavity can be understood geometrically. The dynamics of the turbine are constrained in the present model to its power curve, and thus periodic forcings represent periodic excursions along the power curve centered at some equilibrium point $\lambda_0$. First, let us assume that the local slope at $\lambda_0$ is zero. If the curve is concave-down in a neighborhood about $\lambda_0$, then the value of $C_p$ at $\lambda_0\pm\epsilon$ will be lower than $C_p(\lambda_0)$. Therefore, the integrated value over a periodic excursion away from $\lambda_0$ will be lower than the equilibrium value at $\lambda_0$. The opposite is true when the power curve is concave up; since $C_p(\lambda_0)$ is now a local minimum, nonzero perturbations away from $\lambda_0$ will lead to an increased time-averaged $C_p$. These arguments also hold qualitatively for a nonzero local slope at $\lambda_0$, though the local slope does have an influence on the time-averaged value across a periodic perturbation.

This analysis underscores the point that the geometry of the steady power curve and the equilibrium operating point of the turbine dictate the time-averaged power enhancements or losses that a surging turbine will experience relative to the steady-flow case. In general, it will be favorable to operate in regions of a power curve that have minimal concavity. In terms of turbine design, this implies that turbines whose power curves exhibit a relatively flat maximal region will benefit the most from time-averaged power enhancements relative to the steady case. Flattening the curve in this manner may be achieved through traditional turbine-design methods; alternatively, the pitch of the turbine blades could be varied within a single oscillation period to produce the same topological effect on the $C_p$ manifold. Finally, while concave-up power curves are not typically found in current wind-energy systems, this analysis does suggest that such designs (if physically possible) would lead to even greater power enhancements in unsteady flow conditions.

\subsection{The role of unsteady flow physics}
\label{sec:unsteady}

Until this point, we have neglected the contribution of unsteady fluid mechanics to the system in question. We now investigate these effects analytically. The analysis in this section is purely theoretical, and is included to complete the conceptual picture of wind-energy systems in dynamic-inflow conditions that has been presented in this work. Thus, quantitative predictions and comparisons with the experimental results shown previously are not pursued.

\subsubsection{Specifying velocity potentials for an unsteady extension to 1D momentum theory}

To characterize the influence of streamwise unsteadiness on the theoretical efficiency of wind-energy systems, we extend the analytical framework of \cite{dabiri_theoretical_2020} using the potential-flow modeling approach proposed in Section \ref{sec:model_PDT}. This unsteady extension to the 1D momentum theory of Betz uses the same control volume shown in Figure \ref{fig:CV}, but makes use of two additional unsteady terms:
\begin{equation}
    \frac{d}{dt}\left[KE\right] = \frac{d}{dt}\left[-\frac{1}{2}\rho \iint_{A_2} \phi \hat{n}\cdot \nabla\phi dA\right],
    \label{eqn:KE}
\end{equation}
which represents the unsteady power associated with changes in the streamwise kinetic energy of the actuator disc, and
\begin{equation}
    \Phi_t = \frac{\partial \phi_2}{\partial t} - \frac{\partial \phi_3}{\partial t},
    \label{eqn:Phi_t}
\end{equation}
which is the difference in the unsteady potential across the face of the actuator disc. These quantities are accounted for in energy-conservation relations to obtain expressions for the time-averaged unsteady power. Note that we use $A_i$ to refer to the cross-sectional area at streamwise location $i$; this is not to be confused with the surge-motion amplitude $A$ used elsewhere in this work.

By specifying a velocity potential for the moving actuator disc, we aim to close these two terms and parameterize the instantaneous and time-averaged coefficients of power as a function of the induction factor (as in the steady Betz analysis) and the surge kinematics of the actuator disc. For the sake of simplicity, we will assume that the induction factor $a$ varies in a quasi-steady manner, and will not consider time-derivatives of $a$. Additionally, we only consider the velocity potential associated with the surge velocity $U(t)$, and assume that the velocity potential connected to the free-stream velocity $u_1$ has no unsteady contribution. We begin by completing the analysis for a moving porous disc, and then explore the effects of more general classes of velocity potentials. For the sake of brevity, we omit intermediate steps in the derivations here; several of these details are provided in Appendix \ref{sec:app_unsteady}.

From Equation \ref{eqn:phiPrime}, the velocity potential for a moving porous disc located instantaneously at $x=x_2$ can be written as
\begin{equation}
\phi\left(r,x\right) = a(u_1-U)\sqrt{\frac{2}{\pi}} R^{3/2} \int_0^\infty s^{-1/2}J_{3/2}(Rs)J_0(rs)e^{s (x-x_2)} ds;\; x<x_2,
    \label{eqn:phiDisc}
\end{equation}
and from this expression, the kinetic energy of the disc moving at velocity $U$ can be derived as 
\begin{equation}
    KE_{disc} = \frac{4}{3}\rho a^2 U^2 R^3
\end{equation}
\cite[cf.][Art.\ 102, Eq.\ 20]{lamb_hydrodynamics_1916}. The time derivative is therefore 
\begin{equation}
    \frac{d}{dt}\left[KE_{disc}\right] = \frac{8}{3} \rho R^3 a^2 U \frac{dU}{dt}.
    \label{eqn:dKE_disc}
\end{equation}

The unsteady-potential term $\Phi_t$ can be computed by applying the chain rule to Equation \ref{eqn:phiDisc}, which gives
\begin{equation}
    \frac{\partial \phi_2}{\partial t} = -\frac{2}{\pi}a\sqrt{R^2-r^2}\frac{dU}{dt} + aU^2.
\end{equation}
Using the odd extension to model the region immediately downstream of the disc, we obtain an expression for $\frac{\partial \phi_3}{\partial t}$ that is identical except that the first term in positive. Averaging this over the area of the disc gives
\begin{equation}
    \langle \Phi_t \rangle \equiv \Phi_{t,disc} = -\frac{8R}{3\pi} a\frac{dU}{dt}.
    \label{eqn:Phi_t_disc}
\end{equation}

These results may now be introduced into the framework of \cite{dabiri_theoretical_2020}, with some additional considerations that reduce the parameter space of the theory. The unsteady power associated with the motion of the disc leads to a difference between the available power upstream and downstream of the disc \cite[cf.][Eq.\ 11]{dabiri_theoretical_2020}:
\begin{equation}
    \frac{1}{2}\rho A_2\left(u_2^3-u_3^3\right) = \frac{d}{dt}\left[KE\right].
    \label{eqn:dKE_u2u3}
\end{equation}
Defining additional induction factors $b = 1-u_3/u_1$ and $c = 1-u_4/u_1$, this expression can be written as
\begin{equation}
    (1-a)^3-(1-b)^3=\frac{2}{\rho A_2 u_1^3} \frac{d}{dt}\left[KE\right],
\end{equation}
which we can solve for $b$ in terms of $a$:
\begin{equation}
    b = 1-\left[(1-a)^3-\frac{2}{\rho A_2 u_1^3} \frac{d}{dt}\left[KE\right]\right]^{1/3}.
    \label{eqn:b}
\end{equation}
A nondimensional form of the momentum equation for this problem can be written by means of the unsteady Bernoulli equation as
\begin{equation}
    -2c(1-a) = (1-c)^2 + (1-a)^2 - (1-b)^2 - 1 + \frac{2\Phi_t}{u_1^2},
\end{equation}
which can be solved for the remaining induction factor $c$ as
\begin{equation}
    c = a \pm \sqrt{2a - 1 + (1-b)^2 - \frac{2\Phi_t}{u_1^2}}.
\end{equation}
In this expression, the larger root is taken to satisfy the physical requirement that the flow must slow down between locations 2 and 4, which implies $c > a$. Having written the additional induction factors $b$ and $c$ in terms of the original induction factor $a$ and the known unsteady contributions $\frac{d}{dt}[KE]$ and $\Phi_t$, we can find the instantaneous coefficient of power using the relation
\begin{equation}
    C_p = \frac{1}{2}\left(4c-4c^2+c^3\right) + \frac{1}{2}(2-c)\left[(1-b)^2-(1-a)^2\right] - (2-c)\left(\frac{\Phi_t}{u_1^2}\right).
    \label{eqn:Cp_unsteadyBetz}
\end{equation}
This expression is limited by the physical constraints $b\in[0,1]$, $c\in[0,1]$, and $C_p \geq 0$, which enforce that the velocities at locations 3 and 4 cannot be negative or exceed the free-stream velocity, and that the power extracted from the actuator disc cannot be negative. Note that this analysis does not account for quasi-steady changes in $C_p$ due to the normalization by the effective inflow velocity $u_1 - U(t)$; therefore, changes in $\overline{C_p}$ predicted by this theoretical framework will appear in practice as adjustments to the time-averaged power predictions of the quasi-steady modeling approach outlined in Section \ref{sec:model_pwr}.

\subsubsection{Phase-plane analysis of a surging porous disc}

\begin{figure}
\begin{subfigure}[t]{0.48\textwidth}
\centering
  \includegraphics[width=\textwidth]{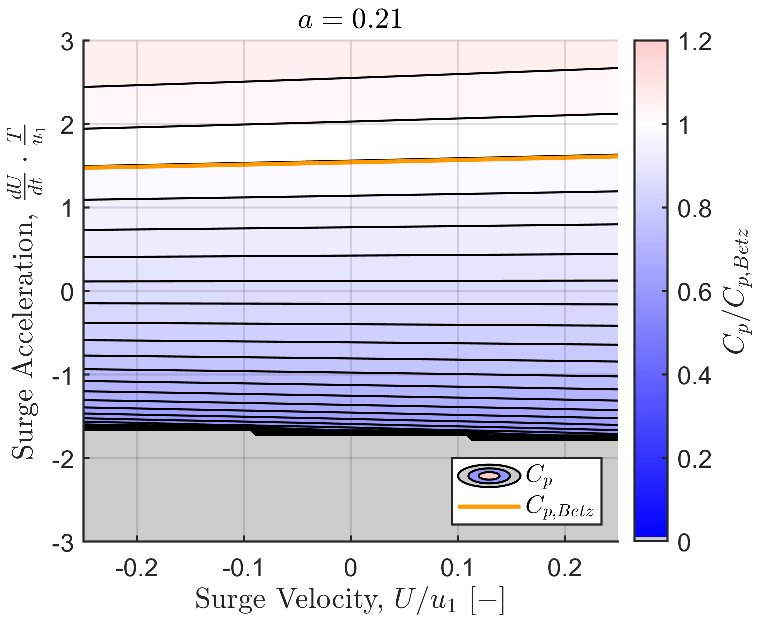}
  \caption{}
\label{fig:Cp_disc_1}
\end{subfigure}
\hfill
\begin{subfigure}[t]{0.48\textwidth}
\centering
  \includegraphics[width=\textwidth]{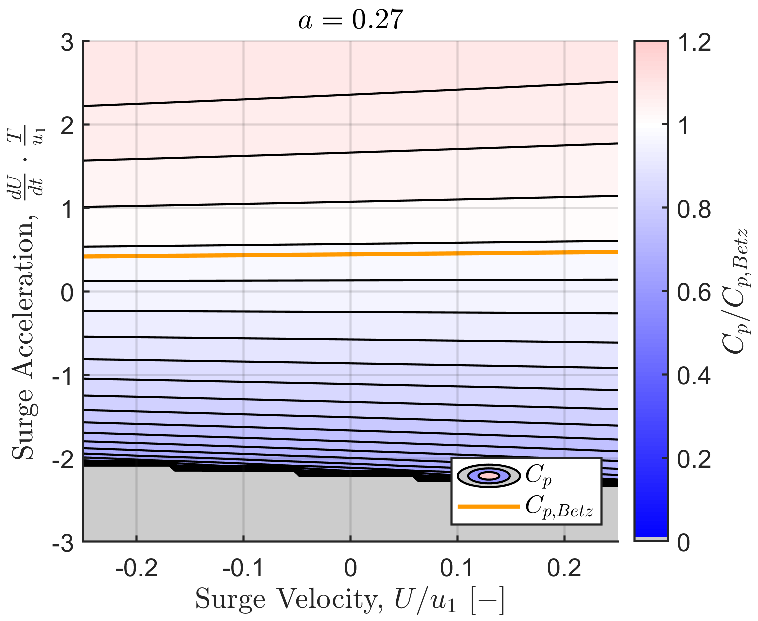}
  \caption{}
\label{fig:Cp_disc_2}
\end{subfigure}
\begin{subfigure}[t]{0.48\textwidth}
\centering
  \includegraphics[width=\textwidth]{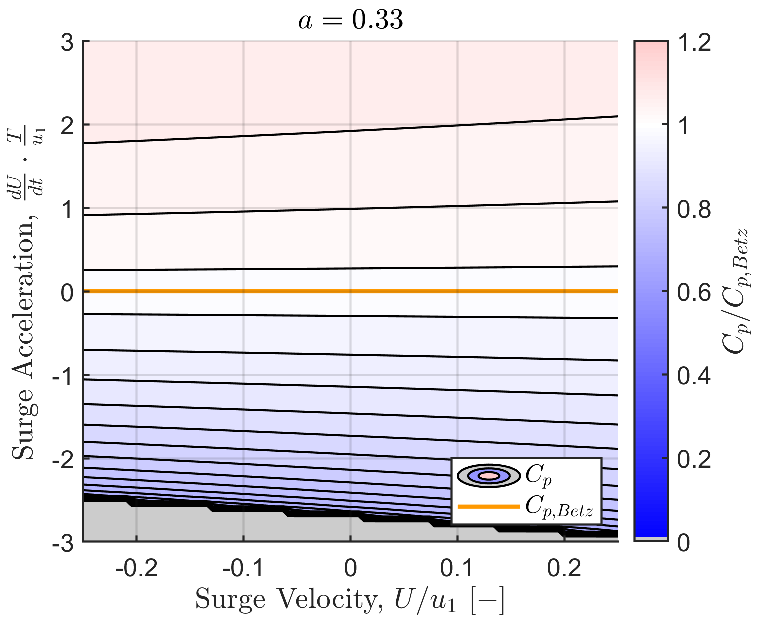}
  \caption{}
\label{fig:Cp_disc_3}
\end{subfigure}
\hfill
\begin{subfigure}[t]{0.48\textwidth}
\centering
  \includegraphics[width=\textwidth]{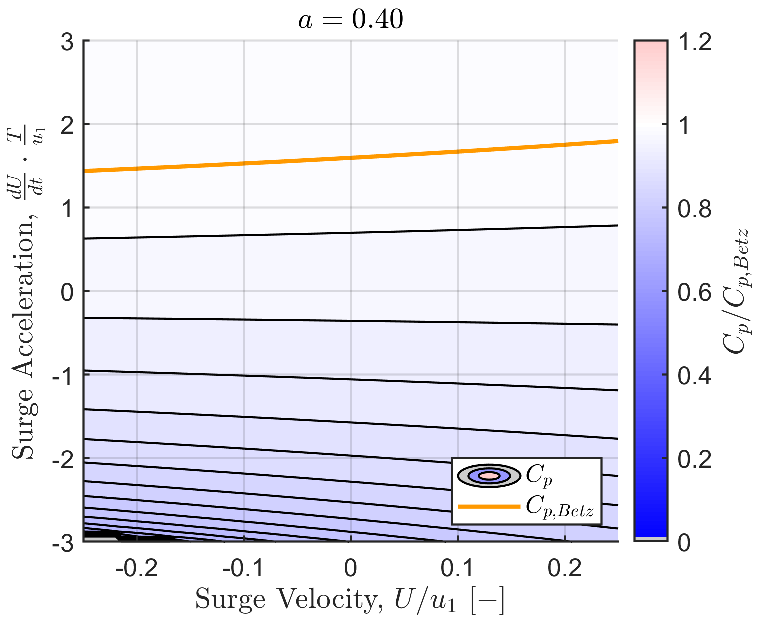}
  \caption{}
\label{fig:Cp_disc_4}
\end{subfigure}
\caption{Contours of $C_p/C_{p,Betz}$ calculated from Equation \ref{eqn:Cp_unsteadyBetz} for a moving porous disc, plotted in the surge-kinematics phase plane as a function of surge velocity and acceleration for induction factors of (a) 0.21, (b) 0.27, (c) $1/3$, and (d) 0.40. The orange line indicates $C_p=C_{p,Betz}$, and grey regions denote dynamics that violate one or more assumptions of the theoretical framework. To nondimensionalize the surge acceleration, $T$ is taken to be 1 s.}
\label{fig:Cp_disc}
\end{figure}

While \cite{dabiri_theoretical_2020} assumed that the parameters $a$, $b$, $c$, and $\Phi_t$ were independent, we now have a theoretical framework that only depends on the induction factor $a$ and the surge kinematics of the actuator disc, $U$ and $\frac{dU}{dt}$. For a given induction factor, we may therefore use the system of equations described above to construct a phase portrait for $C_p$ in terms of the surge kinematics. Contours of $C_p$ in the $U$-$\frac{dU}{dt}$ phase plane are shown for four values of $a$ (0.21, 0.27, $1/3$, and 0.40) in Figure \ref{fig:Cp_disc}. The orange contour marks the Betz efficiency ($C_p/C_{p,Betz}=1$), and grey regions represent locations in the phase plane where one of the physical constraints on $C_p$ is violated.

If $a$ is constant, a surge waveform will appear as a periodic loop in the phase plane that must be centered on the origin so that the turbine has no net displacement. A sinusoidal waveform will follow an elliptical trajectory about the origin, while a trapezoidal waveform will appear as a rectangular trajectory. The time-averaged coefficient of power is calculated by evaluating the line integral of $C_p$ along one closed cycle of this trajectory. The depiction of $C_p$ in the phase plane thus allows the effects of the unsteady velocity potential and power terms from the porous disc to be evaluated by topological reasoning.

To illustrate this line of argumentation, we consider the case where $a=1/3$ (Figure \ref{fig:Cp_disc_3}). A trajectory that is centered on the origin will experience instantaneous values of $C_p$ in excess of the steady-flow Betz limit for positive surge accelerations, while $C_p$ will decrease below $C_{p,Betz}$ for negative surge accelerations. However, the slope of the contours of $C_p$ is greater below the Betz-limit contour than above it. Thus, for a nonzero surge trajectory, the lower $C_p$ values sampled below the zero-acceleration axis will outweigh the higher $C_p$ values sampled above the axis, and the time-averaged coefficient of power will be lower than the steady coefficient of power. This mathematical effect is qualitatively similar to the concavity-based arguments described above in Section \ref{sec:enhancements}.

For values of $a$ that are above or below the steady-flow optimal value of $1/3$, the same arguments hold: the topology of $C_p$ in the phase plane implies that nontrivial surge motions will yield a decrease in time-averaged efficiency as a result of unsteady effects. Furthermore, if $a$ is allowed to vary in a quasi-steady manner, the surge trajectory will exist in a three-dimensional phase space spanned by $U$, $\frac{dU}{dt}$, and $a$. Ascertaining the precise differences in time-averaged power extraction may be done numerically for a given profile of $a(t)$, e.g.\ one obtained from the amplitude and phase data shown in Figure \ref{fig:a_TF}. Still, as the region in the phase plane where efficiencies above the steady Betz limit occur moves away from the origin for $a\neq1/3$, it is apparent that oscillations in $a$ will further decrease the time-averaged efficiency of the system.

\subsubsection{General velocity potentials and the effects of fore-aft asymmetry}

We have shown that the unsteady contribution to the time-averaged efficiency of a wind-energy system modeled as a surging porous disc is negative. However, efficiency enhancements may be attained with a different choice of velocity potential. Consider a general velocity potential for a moving body with translation velocity $\mathbf{U}$, angular velocity $\mathbf{\Omega}$, and circulation $\Gamma$, located instantaneously at a point $\mathbf{x}=\mathbf{x_0}$ in an otherwise quiescent fluid:
\begin{equation}
\phi(\mathbf{x})=\mathbf{U}\cdot\mathbf{\Xi}+\mathbf{\Omega}\cdot\mathbf{\Theta}+\Gamma\left(\frac{\theta}{2\pi}+\psi\right)
\label{eqn:b6-4-10}
\end{equation}
\cite[][Eq.\ 6.4.10]{batchelor_introduction_2000}. Here, boldfaced variables refer to vectors in 3-space. The functions $\mathbf{\Xi}$, $\mathbf{\Theta}$, and $\psi$ are geometric descriptors that only depend on location relative to the body, $\mathbf{x}-\mathbf{x_0}$. For bodies producing zero circulation, $\Gamma = 0$. We also ignore the effect of rotation and set $\mathbf{\Omega}=0$. (The effect of rotation is explained in Appendix \ref{sec:app_unsteady}.) The translational kinetic energy associated with such a moving body scales as $U^2$, and thus the unsteady power term will always scale as
\begin{equation}
    \frac{d}{dt}[KE]\sim U \frac{dU}{dt},
\end{equation}
irrespective of the body geometry parameterized by the function $\mathbf{\Xi}$.

The time derivative of the velocity potential, on the other hand, is affected by body geometry. This can be written as
\begin{equation}
	\frac{\partial \phi}{\partial t} = \frac{d\mathbf{U}}{dt}\cdot \mathbf{\Xi} - \mathbf{U}\cdot\mathbf{u}
\label{eqn:b6-4-22}
\end{equation}
\cite[][Eq.\ 6.4.22]{batchelor_introduction_2000}, where $\mathbf{u}$ is the local flow velocity at $\mathbf{x}$. To compute $\Phi_t$ from the unsteady theoretical framework, we note that the local flow velocity in the quiescent-fluid frame at streamwise locations 2 and 3 scales with the velocity of the body, $U$. Additionally, since the velocity on either side of the translating body must decrease with increasing distance from the body, the geometric function $\mathbf{\Xi}$ must exhibit odd symmetry about the body plane.

For a translating symmetric body in potential flow, $\mathbf{u}$ on either side of the body will be identical. Therefore, when taking the difference of the unsteady-potential terms upstream and downstream of the body, the $\mathbf{U}\cdot\mathbf{u}$ term will cancel. The odd symmetry of the geometric function $\mathbf{\Xi}$ will retain the acceleration-dependent term, and thus for a symmetric body, we have
\begin{equation}
    \Phi_t \sim \frac{dU}{dt}.
\end{equation}

For a body that exhibits fore-aft asymmetry about its flow-normal center plane, however, $\mathbf{u}$ may differ across the upstream and downstream sides of the body. In this case, the quadratic velocity term may not cancel, suggesting that
\begin{equation}
    \Phi_t \sim R\frac{dU}{dt} \pm U^2.
\end{equation}
The additional dependence on $U^2$ will change the topology of $C_p$ in the phase plane and affect the time-averaged coefficients of power that are possible in the theoretical framework.

\begin{figure}
\begin{subfigure}[t]{0.48\textwidth}
\centering
  \includegraphics[width=\textwidth]{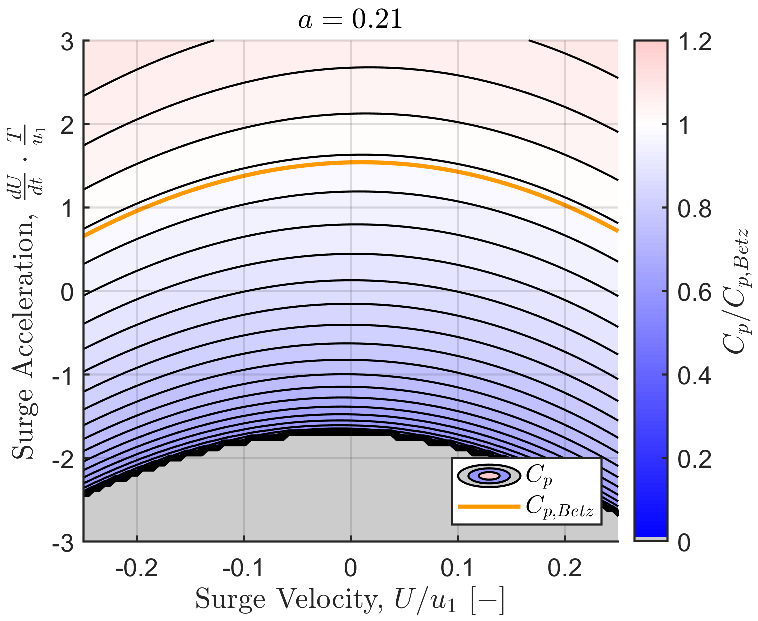}
  \caption{}
\label{fig:Cp_asym_1}
\end{subfigure}
\hfill
\begin{subfigure}[t]{0.48\textwidth}
\centering
  \includegraphics[width=\textwidth]{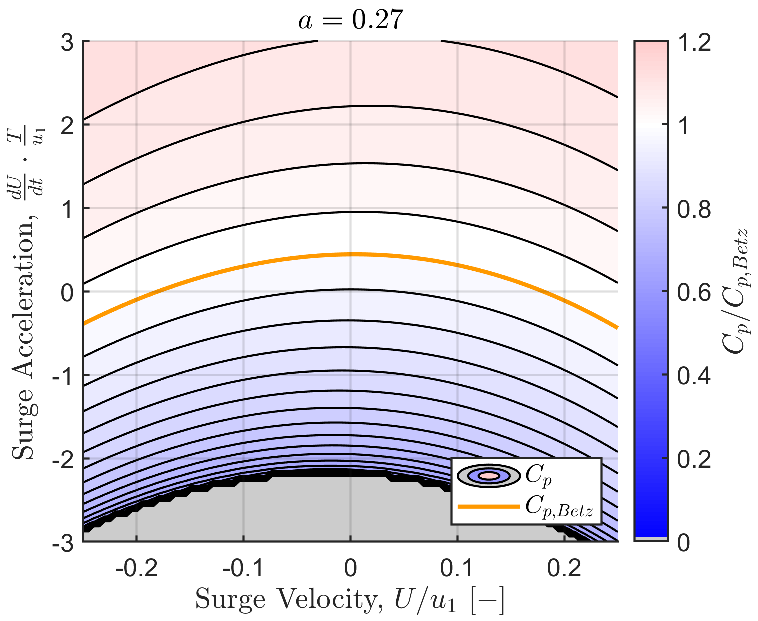}
  \caption{}
\label{fig:Cp_asym_2}
\end{subfigure}
\begin{subfigure}[t]{0.48\textwidth}
\centering
  \includegraphics[width=\textwidth]{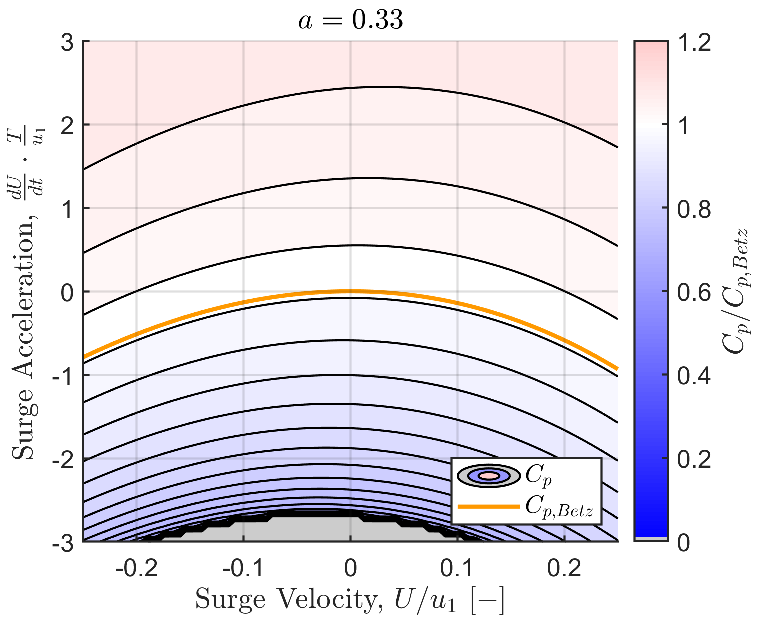}
  \caption{}
\label{fig:Cp_asym_3}
\end{subfigure}
\hfill
\begin{subfigure}[t]{0.48\textwidth}
\centering
  \includegraphics[width=\textwidth]{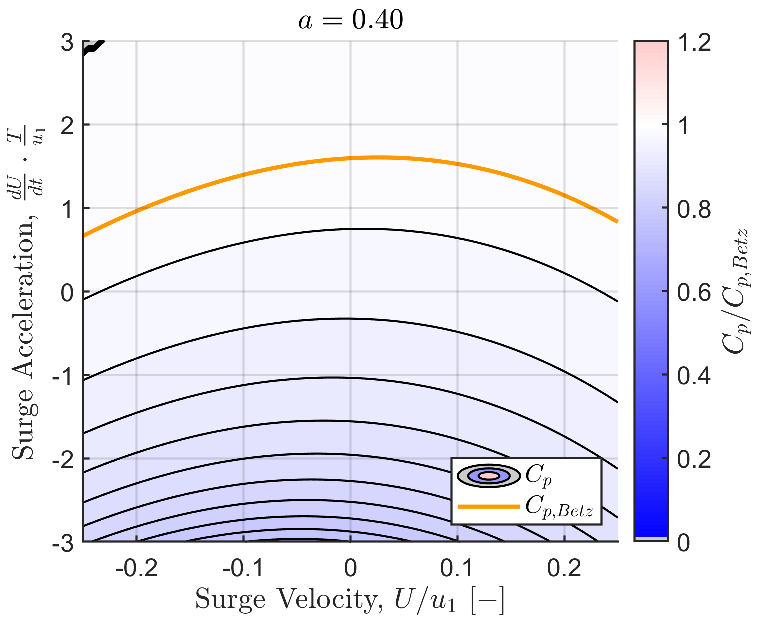}
  \caption{}
\label{fig:Cp_asym_4}
\end{subfigure}
\caption{Contours of $C_p/C_{p,Betz}$ for a moving porous asymmetric body (parameterized by Equation \ref{eqn:Phi_t_asym}), plotted in the surge-kinematics phase plane as a function of surge velocity and acceleration for induction factors of (a) 0.21, (b) 0.27, (c) $1/3$, and (d) 0.40. The orange line indicates $C_p=C_{p,Betz}$, and grey regions denote dynamics that violate one or more assumptions of the theoretical framework.}
\label{fig:Cp_asym}
\end{figure}

To illustrate these effects, we conduct a phase-plane analysis by assuming the existence of a velocity potential for a moving asymmetric body that yields
\begin{equation}
    \Phi_t =  -\frac{8}{3\pi}a\left( R\frac{dU}{dt} + U^2\right).
    \label{eqn:Phi_t_asym}
\end{equation}
We keep the same scaling coefficients as for the surging porous disc for the sake of comparison, and use the same expression for the kinetic energy of the body (Equation \ref{eqn:dKE_disc}). Contours of $C_p$ for the same four values of $a$ shown previously are given for this representative model in Figure \ref{fig:Cp_asym}. We again first consider the case where $a=1/3$ (Figure \ref{fig:Cp_asym_3}). Here, we observe that the addition of the $U^2$ term in the expression for $\Phi_t$ has curved the contours of $C_p$ such that the region where $C_p>C_{p,Betz}$ now extends below the zero-acceleration axis. This implies that a trajectory centered on the origin can have a time-averaged efficiency that exceeds the steady Betz limit. This may also be possible if we allow for small oscillations in $a$, depending on the surge kinematics applied and their corresponding trajectories in the phase space.

It is important to note that if the sign of the $U^2$ term in Equation \ref{eqn:Phi_t_asym} were reversed, the concavity of the contours would also be reversed, leaving the region where $C_p>C_{p,Betz}$ on the interior of a concave-up parabola with respect to the surge velocity. In this scenario, a periodic trajectory will spend less time in the efficiency-enhancing region relative to the efficiency-depleting region, and the time-averaged efficiency will therefore drop below the steady Betz limit. Thus, fore-aft asymmetry in the velocity potential is a necessary but not sufficient condition for enhancements in the time-averaged efficiency of a periodically translating actuator body.

This conceptual exercise demonstrates that, for symmetric bodies in potential flow, the unsteady contribution to time-averaged efficiency will be negative. As the modeling efforts and experiments presented previously seem to suggest that a moving porous-disc model captures the dominant dynamics of a periodically surging horizontal-axis wind turbine, it is likely that unsteady effects would act against the time-averaged power enhancements described in Section \ref{sec:enhancements} above. This hypothesis appears to be consistent with the time-averaged power data shown in Figure \ref{fig:Pwr_vs_U}, which at high surge-velocity amplitudes tend to be overpredicted by the nonlinear turbine model (which assumes quasi-steady aerodynamics). However, wind-energy systems need not be symmetric. Streamwise asymmetries across the rotor plane could be introduced either mechanically, through the geometry of the turbine, or dynamically, through intracycle blade-pitch or generator-load control. These non-traditional design and control paradigms could create beneficial asymmetries in the equivalent velocity potential of the system that could be leveraged to achieve higher time-averaged efficiencies than steady-flow or quasi-steady analyses would suggest. In the case of dynamic induction-control schemes, the influence of unsteady induction-factor variations $\frac{da}{dt}$ may no longer be negligible, and the analysis presented here would need to be expanded to include these variations in a four-dimensional phase space. Still, the present analysis can serve as a helpful theoretical framework for characterizing trends in unsteady contributions to the efficiency of wind-energy systems.

\section{Conclusions}
\label{sec:conclusions}

In this work, a nonlinear dynamical model for the power generation of a periodically surging wind turbine was paired with a potential-flow model for the flow properties upstream of the turbine. This modeling framework was shown to reproduce trends in experimental measurements of both the time-averaged power extraction and upstream flow-velocity and pressure, at surge-velocity amplitudes of up to 24\% of the wind speed. These results are posited to be equally applicable to stationary turbines in dynamically varying inflow conditions, such as axial gusts. A key advantage of this approach is that the entire modeling approach is calibrated only by steady-flow quantities: the turbine power curve and the radial induction profile of the turbine at the rotor plane. The theoretical analyses also identified and parameterized contributions to power-extraction enhancements over the steady-flow case, such as a dependence on the local concavity of the turbine power curve and the role of streamwise asymmetries in unsteady power gains. This work thus not only informs the design, characterization, and control of wind and hydrokinetic turbines in unsteady flow environments, such as floating offshore wind farms and tidal currents, but also yields fundamental insights into the relative influences of quasi-steady and unsteady fluid mechanics in energy-harvesting systems.

While similar theoretical tools have been widely applied to the analysis of wind turbines in steady flow, a major contribution of this work is the extension of these methods to unsteady flow contexts. The porous-disc model is also similar in principle to the actuator-disc models often used in numerical simulations of large wind farms \cite[e.g.][]{calaf_large_2010,stevens_flow_2017}, and thus this study could inform modifications of existing actuator-disc simulations for surging-turbine or dynamic-inflow conditions. This may be particularly useful for large-eddy simulations (LES) of floating offshore turbine arrays, where the analytical turbine model can help to parameterize the coupling between turbine inflow conditions, sea-surface waves, and floating-platform dynamics. Additionally, the induction and thrust-force predictions of this modeling framework could be used as initial conditions for wake models of turbines in dynamic inflow conditions, which could further improve parameterizations of turbine aerodynamics in numerical simulations. Such a connection was recently developed for yawed turbines by \cite{heck_modelling_2023}, whose modeling philosophy was influential in the development of the present analytical approach. These wake-modeling initiatives would further benefit from the work of \cite{steiros_drag_2018} and \cite{bempedelis_analytical_2022}, particularly for capturing wake-pressure effects in highly loaded turbine configurations.

This work has several implications for full-scale wind-energy systems in real-world flow conditions. First, the analytical model for flow properties upstream of a surging turbine can be used in conjunction with nacelle-mounted LiDAR units for improved load control and wind-speed estimation in floating offshore applications. The same principles can be applied to stationary turbines in gusty environments and kite-mounted aerial turbines. Secondly, these analytical and experimental results reinforce and parameterize the evidence collected by \cite{wen_influences_2017}, \cite{el_makdah_influence_2019}, \cite{johlas_floating_2021}, and \cite{wei_phase-averaged_2022} that streamwise unsteadiness (either in the flow or in the turbine itself) can lead to increases in power extraction above the reference steady case. The present investigations suggest both quasi-steady and unsteady mechanisms that can be exploited to capitalize on these power-extraction enhancements. Future work can investigate turbine design and control strategies, such as active blade pitching and intracycle load control \citep[e.g.][]{strom_intracycle_2017}, that may further increase the time-averaged power extraction of floating offshore wind turbines and other systems that can operate in inherently unsteady flows. The modeling framework also provides a means to estimate thrust loads on turbines from dynamic inflow conditions, which may increase fatigue loading on turbine blades and support structures. The analytical tools outlined in this study can inform control strategies that anticipate the changes in thrust and blade loading as a function of changing inflow conditions and dynamically adjust the blade pitch or generator load to mitigate the unsteady loads of unwanted disturbances and oscillations. Whether these models are used to enhance unsteady power-conversion gains or extend the operational lifespan of energy-harvesting systems by controlling unsteady loads, this work makes the case that unsteady flow phenomena should be at the forefront of design considerations for structures operating in the atmosphere and ocean.

\backsection[Acknowledgements]{The authors would like to thank Michael Howland for several enlightening conversations regarding the development of the nonlinear dynamical model and its connection to induction effects. The authors also would like to thank Konstantinos Steiros for sharing resources that helped with the derivation of the porous-disc model, and the three anonymous referees assigned by the \textit{Journal of Fluid Mechanics} for their feedback and suggestions, which greatly strengthened the analysis and conclusions of this study. The authors additionally appreciate the assistance and safety supervision of Malaika Cordeiro, Matthew Fu, and Peter Gunnarson during the experiments.}

\backsection[Funding]{This work was supported by the National Science Foundation (grant number CBET-2038071) and the Caltech Center for Autonomous Systems and Technologies. Nathaniel Wei was supported by a National Science Foundation Graduate Research Fellowship.}

\backsection[Declaration of interests]{The authors report no conflict of interest.}

\backsection[Data availability statement]{The data that support the findings of this study are available upon request.}

\backsection[Author ORCID]{N.\ J.\ Wei, https://orcid.org/0000-0001-5846-6485; J.\ O.\ Dabiri, https://orcid.org/0000-0002-6722-9008}


\appendix{}

\section{Vortex-cylinder theory results}
\label{sec:app_VCT}

\begin{figure}
\begin{subfigure}[t]{0.48\textwidth}
\centering
  \includegraphics[width=\textwidth]{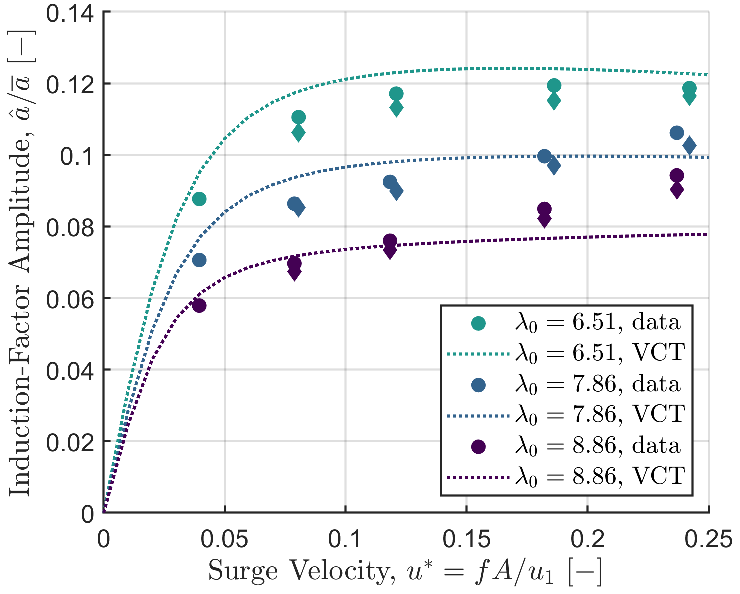}
  \caption{}
\label{fig:a_amp_VCT}
\end{subfigure}
\hfill
\begin{subfigure}[t]{0.48\textwidth}
\centering
  \includegraphics[width=\textwidth]{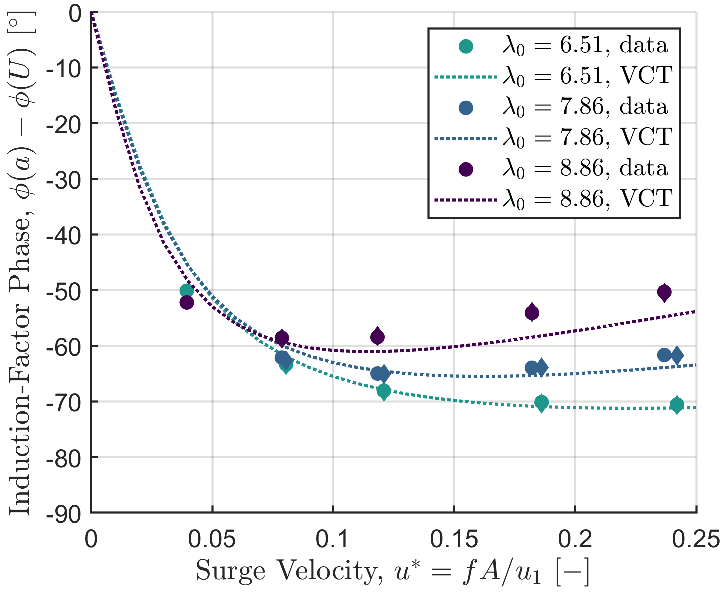}
  \caption{}
\label{fig:a_phase_VCT}
\end{subfigure}
\caption{(a) Amplitude and (b) phase of the estimated induction factors using the VCT model, plotted against surge-velocity amplitude. Model predictions are given as dotted lines.}
\label{fig:a_TF_VCT}
\end{figure}

\begin{figure}
\begin{subfigure}[t]{0.48\textwidth}
\centering
  \includegraphics[width=\textwidth]{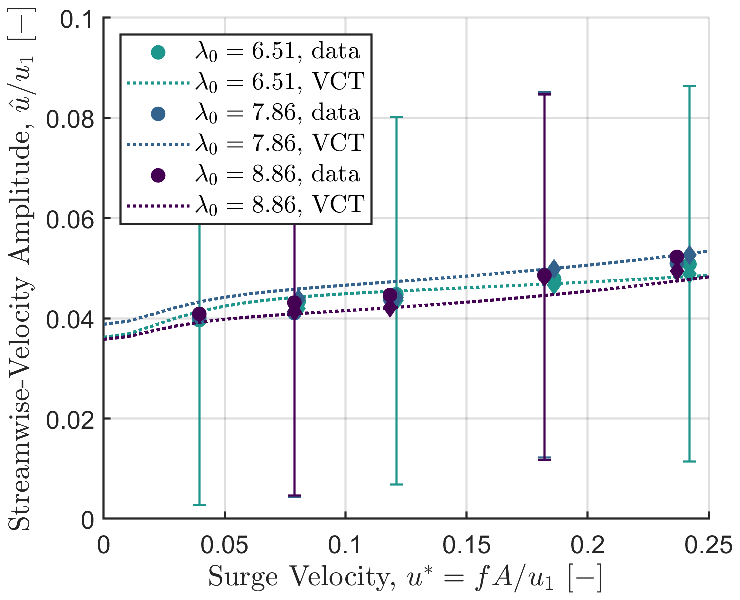}
  \caption{}
\label{fig:uHW_amp_VCT}
\end{subfigure}
\hfill
\begin{subfigure}[t]{0.48\textwidth}
\centering
  \includegraphics[width=\textwidth]{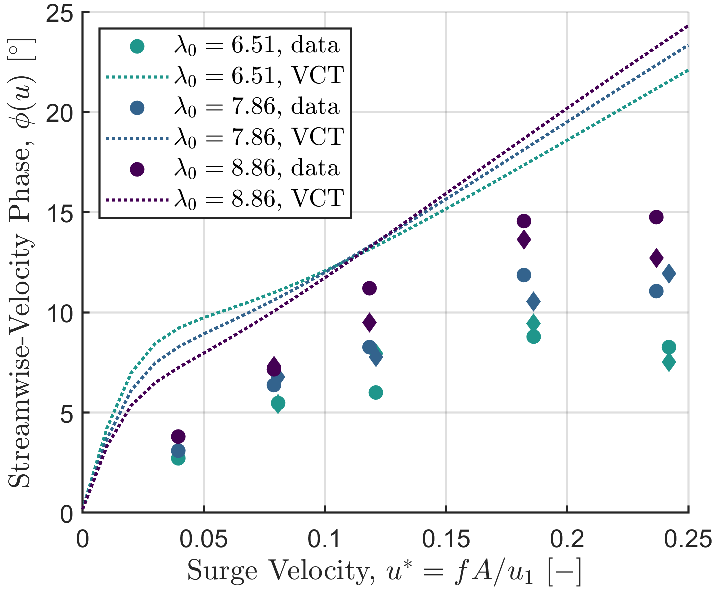}
  \caption{}
\label{fig:uHW_phase_VCT}
\end{subfigure}
\caption{(a) Amplitude and (b) phase of the measured flow velocity at $x=x_u$, plotted against surge-velocity amplitude. VCT model predictions are given as dotted lines. Error bars are plotted on every fourth point.}
\label{fig:uHW_TF_VCT}
\end{figure}

\begin{figure}
\begin{subfigure}[t]{0.48\textwidth}
\centering
  \includegraphics[width=\textwidth]{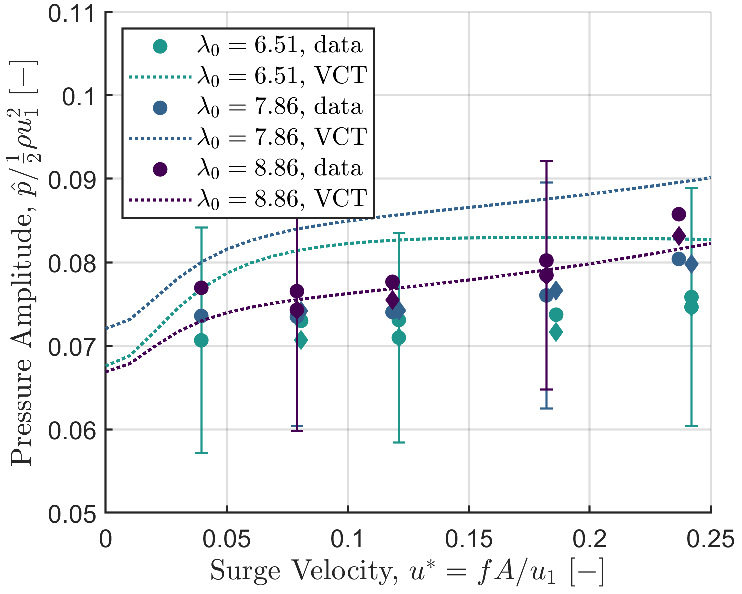}
  \caption{}
\label{fig:p_amp_VCT}
\end{subfigure}
\hfill
\begin{subfigure}[t]{0.48\textwidth}
\centering
  \includegraphics[width=\textwidth]{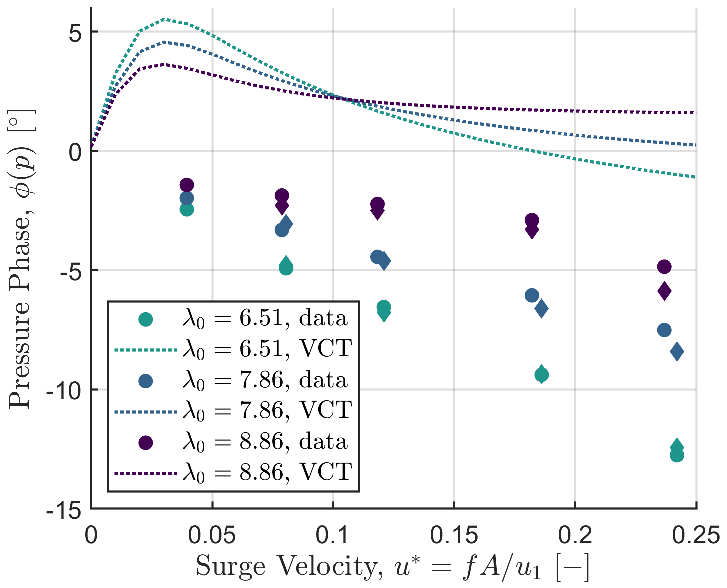}
  \caption{}
\label{fig:p_phase_VCT}
\end{subfigure}
\caption{(a) Amplitude and (b) phase of the measured relative pressure at $x=x_u$, plotted against surge-velocity amplitude. VCT model predictions are given as dotted lines. Error bars are plotted on every fourth point.}
\label{fig:p_TF_VCT}
\end{figure}

Since Figures \ref{fig:a_TF}, \ref{fig:uHW_TF}, and \ref{fig:p_TF} in Section \ref{sec:results_flow} omitted the predictions of VCT, these results are included in this appendix. The data in Figures \ref{fig:a_TF_VCT}, \ref{fig:uHW_TF_VCT}, and \ref{fig:p_TF_VCT} are identical to the aforementioned figures, but VCT model predictions are shown instead of PDT predictions. Similar trends are observed, though slight quantitative differences exist between the two sets of model predictions. Overall, both models appear to capture the trends observed in the data reasonably well.

\FloatBarrier

\section{Derivations from the unsteady theoretical analysis}
\label{sec:app_unsteady}

\subsection{Porous-disc derivations}

The kinetic energy associated with the motion of a solid disc is given by
	
	\begin{equation}
	KE_{disc} = -\frac{1}{2}\rho \iint_A \phi \hat{n}\cdot \nabla \phi dA,
	\label{eqn:KE_disc}
	\end{equation}
	
	\noindent where the flux integral is taken on an infinitesimally thin control surface surrounding the disc. Across this surface, $\nabla \phi = U\hat{i}$. The kinetic energy of the disc can thus be written as
	
	\begin{equation}
	\begin{split}
	KE_{disc} &= -\frac{1}{2}\rho U \iint_A \phi \left(\hat{n}\cdot\hat{i}\right) dA\\
	&= -\frac{1}{2}\rho U \left[-\int_0^R 4U\sqrt{R^2-r^2}rdr-\int_0^R 4U\sqrt{R^2-r^2}rdr\right]\\
	&= 4\rho U^2 \int_0^R r\sqrt{R^2-r^2} dr\\
	&= \frac{4}{3}\rho U^2 R^3.
	\end{split}
	\label{eqn:KE_disc2}
	\end{equation}

 The odd extension to the upstream velocity potential given in Equation \ref{eqn:phiDisc} is
 \begin{equation}
\phi\left(r,x\right) = -a(u_1-U)\sqrt{\frac{2}{\pi}} R^{3/2} \int_0^\infty s^{-1/2}J_{3/2}(Rs)J_0(rs)e^{-s (x-x_2)} ds;\; x>x_2.
    \label{eqn:phiDiscDownstream}
\end{equation}

 We may differentiate the expression for $\phi$ at streamwise location 3 in time, noting that $x_3$ represents a fixed spatial coordinate with zero time derivative, while $x_2$ represents the instantaneous position of the actuator disc such that $\frac{dx_2}{dt}=U$:
	
	\begin{multline}
	\begin{split}
	\frac{\partial \phi_3}{\partial t} &= a\frac{d U}{d t} \left[R\left(\frac{2R}{\pi}\right)^{1/2}\int_0^{\infty} s^{-1/2} e^{-s\left(x_3-x_2\right)}J_0(rs)J_{3/2}(Rs)ds\right] \\
	&\;\;+ aU\frac{\partial}{\partial t}\left[R\left(\frac{2R}{\pi}\right)^{1/2}\int_0^{\infty} s^{-1/2} e^{-s\left(x_3-x_2\right)}J_0(rs)J_{3/2}(Rs)ds\right]\\
	&= a\frac{2}{\pi}\sqrt{R^2-r^2}\frac{dU}{dt} + aUR\left(\frac{2R}{\pi}\right)^{1/2}\int_0^{\infty} s^{-1/2} \frac{d}{dt}\left[e^{-s\left(x_3-x_2\right)}\right]J_0(rs)J_{3/2}(Rs)ds\\
	&= a\frac{2}{\pi}\sqrt{R^2-r^2}\frac{dU}{dt} + aUR\left(\frac{2R}{\pi}\right)^{1/2}\int_0^{\infty} s^{-1/2} \left[s \frac{dx_2}{dt}e^{-s\left(x_3-x_2\right)}\right]J_0(rs)J_{3/2}(Rs)ds\\
	&= a\frac{2}{\pi}\sqrt{R^2-r^2}\frac{dU}{dt} + aUR\left(\frac{2R}{\pi}\right)^{1/2}\int_0^{\infty} s^{1/2} Ue^{-s\left(x_3-x_2\right)}J_0(rs)J_{3/2}(Rs)ds\\
	&= a\frac{2}{\pi}\sqrt{R^2-r^2}\frac{dU}{dt} + aU^2.
	\end{split}
	\label{eqn:dphidt3}
	\end{multline}
The following integrals are used in this derivation:
\begin{equation}
	\int_0^\infty s^{-1/2}J_0(rs)J_{3/2}(Rs) ds = \sqrt{\frac{2}{\pi}} \sqrt{R^2-r^2} R^{-3/2},\; 0\leq r \leq R
	\label{eqn:int2}
	\end{equation}
	and
	\begin{equation}
	\int_0^\infty s^{1/2}J_0(rs)J_{3/2}(Rs) ds = \sqrt{\frac{\pi}{2}}R^{-3/2},\; 0\leq r \leq R.
	\label{eqn:int3}
	\end{equation}
 By a similar process, we obtain
 \begin{equation}
	\frac{\partial \phi_2}{\partial t} = -a\frac{2}{\pi}\sqrt{R^2-r^2}\frac{dU}{dt} + aU^2.
	\label{eqn:dphidt2}
	\end{equation}
Combining these two unsteady potentials, we arrive at the relation
\begin{equation}
    \Phi_t = -\frac{4}{\pi}a\sqrt{R^2-r^2}\frac{dU}{dt},
\end{equation}
which may be averaged across the face of the actuator disc to obtain
\begin{equation}
	\langle\Phi_t\rangle = \frac{2\pi}{A_2} \int_0^R \left[-\frac{4}{\pi}a\sqrt{R^2-r^2} \frac{dU}{dt}\right]rdr=-\frac{8R}{3\pi}a\frac{dU}{dt}\equiv \Phi_t.
\end{equation}

\subsection{Rotating actuator bodies}
	
	We consider the case of a purely rotating actuator body, for which $\mathbf{U}=0$ and $\mathbf{\Omega}\neq0$. Equation \ref{eqn:b6-4-10} becomes
	
	\begin{equation}
	\phi(\mathbf{x}) = \mathbf{\Omega}\cdot\mathbf{\Theta},
	\label{eqn:phi_rot}
	\end{equation}
	
	\noindent where $\mathbf{\Omega}$ is an axial vector. The kinetic energy of the rotating body scales as $\left|\mathbf{\Omega}\right|^2$, so that in the quasi-1D formulation, we find
	
	\begin{equation}
	\frac{d}{dt}(KE) \sim \Omega \frac{d\Omega}{dt}, 
	\label{eqn:dKE_rot}
	\end{equation}
	
	\noindent ignoring geometric constants. This is analogous to the case of a purely translating body, for which the time derivative of the kinetic energy scales as $U \frac{dU}{dt}$.
	
	Taking the time derivative of Equation \ref{eqn:phi_rot} yields
	
	\begin{equation}
	\frac{\partial \phi}{\partial t} = \frac{\partial\mathbf{\Omega}}{\partial t}\cdot \mathbf{\Theta} + \mathbf{\Omega}\cdot\frac{\partial\mathbf{\Theta}}{\partial t} = \frac{\partial\mathbf{\Omega}}{\partial t}\cdot \mathbf{\Theta} + \mathbf{\Omega}\cdot \left(\frac{\partial(\mathbf{x}-\mathbf{x_0})}{\partial t}\cdot\nabla\mathbf{\Theta}\right).
	\end{equation}
	
	Noting that $\frac{\partial(\mathbf{x}-\mathbf{x_0})}{\partial t} = -\mathbf{U} = 0$, the second term vanishes. For the quasi-1D case, we write the vector-valued functions $\mathbf{\Omega}$ and $\mathbf{\Theta}$ as scalar-valued functions and obtain the result
	
	\begin{equation}
	\frac{\partial\phi}{\partial t} = \Theta\frac{d\Omega}{dt}.
	\label{eqn:Phit_rot}
	\end{equation}
	
	Comparing Equations \ref{eqn:dKE_rot} and \ref{eqn:Phit_rot} with the previously obtained results for a purely translating body with a locally symmetric flow field, we see that there is no symmetry-breaking term for a purely rotating body that will yield time-averaged efficiencies in excess of the Betz limit. This aligns with intuition, as changes in the rotation rate of an actuator body should affect the flow on either side of the body symmetrically, which, as we have seen, will not lead to net improvements in efficiency. It is possible that, in a real wind turbine, changes in the rotation rate would lead to unsteadiness at the turbine-blade level, potentially due to dynamic stall or other kinds of vortex-shedding events. These effects, however, would manifest themselves as streamwise unsteadiness (e.g.\ changes in induced velocities) rather than rotational unsteadiness. Thus, streamwise unsteadiness remains the primary parameter of interest for these investigations.

\bibliographystyle{jfm}
\bibliography{Induction}

\end{document}